\documentclass[12pt,preprint]{aastex}
\def\kms{{\rm km\,s^{-1}}}
\def\pc{{\rm pc}}

\def\masyr{{\rm mas}\,{\rm yr}^{-1}}
\def\usno{{\rm USNO}}
\def\nltt{{\rm NLTT}}
\def\rpm{{\rm RPM}}
\def\lim{{\rm lim}}

\begin{document}

\title{Improved Astrometry and Photometry for the Luyten Catalog. 
II. Faint Stars and the Revised Catalog}

\author{Samir Salim and Andrew Gould}
\affil{Department of Astronomy, The Ohio State University,
140 W.\ 18th Ave., Columbus, OH 43210}
\email{samir,gould@astronomy.ohio-state.edu}

\singlespace

\begin{abstract}

	We complete construction of a catalog containing improved astrometry
and new optical/infrared photometry for the vast majority of NLTT stars lying 
in the overlap of regions covered by POSS I and by the second 
incremental 2MASS release, 
approximately 44\% of the sky.  The epoch 2000 positions are typically accurate
to 130 mas, the proper motions to $5.5\,\masyr$, and the $V-J$ colors to 
0.25 mag.  Relative proper motions of binary components are meaured to 
$3\,\masyr$.  The false identification rate is $\sim 1\%$ for $11\la V\la 18$
and substantially less at brighter magnitudes.
These improvements permit the construction of a
reduced proper motion diagram that, for the first time, allows one to
classify NLTT stars into main-sequence (MS) stars, subdwarfs (SDs), and 
white dwarfs (WDs).
We in turn use this diagram to analyze the properties of both our catalog
and the NLTT catalog on which it is based.  In sharp contrast to popular
belief, we find that NLTT incompleteness in the plane is almost completely
concentrated in MS stars, and that SDs and WDs are detected almost 
uniformly over the sky $\delta>-33^\circ$.  Our catalog will therefore
provide a powerful tool to probe these populations statistically, as
well as to reliably identify individual SDs and WDs.

\end{abstract}
\keywords{astrometry --  catalogs -- methods: statistical}
\clearpage
 
\section{Introduction
\label{sec:intro}}

	In Paper I \citep{paper1}, we discussed the motivation for 
improving the astrometry and the photometry of the 
$\sim 59,000$ high proper-motion ($\mu\geq 180\,\masyr$) stars in 
the {\it New Luyten Two Tenths} (NLTT) Catalog \citep{luy}, and we 
outlined our basic strategy for obtaining these improvements:
at the bright end, match NLTT star to entries in the
Hipparcos \citep{hip}, Tycho-2 \citep{t2}, and Starnet \citep{starnet} catalogs
and, at the faint end, find counterparts of
USNO-A \citep{usnoa1,usnoa2} stars in 2MASS \citep{2mass} whose position 
offsets are predicted by the proper motions listed in NLTT.  We carried out
the bright-end search and used the results to characterize the
position and proper-motion (PPM) errors of the NLTT.

	Here we complete the catalog (in the regions covered by the first 
2MASS release and the first Palomar Observatory Sky Survey -- POSS I) 
by carrying out
the faint-end search, and we integrate the results of both searches.
This paper therefore has three interrelated goals. First, to give a 
comprehensive account of how the catalog was constructed and of the 
various tests we performed to determine the accuracy of our 
identifications and our proper-motion measurements.  
Second, to characterize the properties of the original 
NLTT catalog, now that much better astrometry and photometry are available
for a large fraction of its stars.  Third, to characterize the properties
of the catalog we have constructed, including its precision and its 
completeness relative to NLTT.  We expect that the importance of the
first and third goals are obvious to the reader, but the reasons for
characterizing the NLTT just at the time when it is being superseded may not
be.

	First, in constructing the present catalog, we made heavy use of
all aspects of NLTT including not only its positions, proper motions,
and photometry, but also its extensive notes on binaries.  Our approach
to carrying out this work was heavily influenced by our assessment of NLTT's
properties, and hence ours catalog's construction and limitations cannot be 
fully understood without a knowledge of these properties.  Second, the
publication of NLTT was a watershed in the history of astronomy: the
lengthy list of papers that we cited in Paper I comprise
but a tiny fraction of the literature that is based directly or indirectly
on NLTT.  A good understanding of the NLTT's properties will be useful
in assessing which of the conclusions drawn by these papers can be relied
upon, and which require further refinement.  Third, more work revising the
NLTT remains to be done.  For technical reasons that will be described below,
our catalog will cover only the portion of the sky $\delta > -33^\circ$.
And, of course, as of now it is restricted to the $\sim 1/2$ of sky covered
by the 2MASS second incremental release.  
Hence, for the present, the original NLTT remains
the best source of information for 23,000 of its 59,000 stars.  

\section{Strategy for Matching NLTT to USNO-A and 2MASS
\label{sec:strategy}}

\subsection{Error Characterization
\label{sec:err_char}}

	In Paper I, we showed that, at least at the bright end,
NLTT position errors have several characteristic scales.  For
the majority of stars, the positions are measured to a precision
of $1''$, although they are recorded only to 1 s of time in RA
and $6''$ of arc in DEC\footnote{About $10\%$ are recorded to
worse precision: 6 s of time and $1'$ of arc.  See below.}.
For most of the remainder, the
intrinsic measurement errors are $\sim 6''$ although, again,
the errors in the tabulated values are somewhat increased
by decimal truncation.  However, there is also a ``halo''
of position errors extending out to several arcmin, as well
as a handful of outliers with much greater errors.  The proper
motion errors were found to be ``correlated'' with the position errors in
the sense that the stars with the better position measurements 
also have better proper motion errors ($\sim 25\,\masyr$), while
those with worse positions have somewhat larger proper motion
errors ($\sim 35\,\masyr$).  These characteristics motivated
a basically two-tiered approach to matching bright NLTT stars with
the PPM catalogs, and it will also motivate a two-tiered approach
here.

\subsection{Scope of the Faint Search
\label{sec:scope}}

	As mentioned in Paper I, we wish to push our ``faint''
search as bright as possible, thereby maximizing overlap with
the bright search.  While we expect that the overwhelming
majority of the bright-search matches between NLTT and the 
three PPM catalogs (Hipparcos, Tycho-2, Starnet) are genuine, 
there is an important path to false matches.  The true 
counterpart of the NLTT star may not be in any of the three 
PPM catalogs.  Then, even though the real star may (most likely 
does) have a PPM that is very
close to its NLTT values, our bright-star approach is to search
farther and farther from the NLTT PPM in these catalogs.
Eventually, we may find a barely acceptable match in these
catalogs.  Such cases will be easily uncovered by cross-checking
the bright and faint searches.

	However, we cannot push our faint-star approach of
matching counterparts from USNO-A and 2MASS too bright because
bright stars are increasingly saturated in the POSS plates that 
were scanned to construct USNO-A, leading to increasingly
unreliable PPMs and even identifications.

We therefore begin the ``faint'' search
by removing from consideration all stars that 
were matched to Hipparcos in the first two (``rectangle'' and 
``circle'') searches carried out in Paper I.  The Hipparcos completeness
limit $(V=7.3)$ assures us that essentially all the very bright stars in
NLTT are accounted for in this way.  Moreover, the high-proper 
motion stars that entered the Hipparcos input catalog beyond its 
magnitude
limit were drawn almost entirely from NLTT.  (This is confirmed
by the fact that in Paper I, essentially all Hipparcos high
proper-motion stars were successfully matched to NLTT.  By 
contrast, for Tycho-2, whose magnitude range strongly overlaps
that of Hipparcos but was not constructed from an input catalog, 
there were several hundred non-matches.)\ \ 
Hence, the probability that the ``best'' match of a Hipparcos star 
to NLTT is a false match is very low.  However, from the faint 
search we do not exclude Hipparcos stars that were matched using
our more aggressive strategies, in order to have an additional check on the
robustness of these identifications.

Next, we restrict our search to portions of the sky covered
by the (circa 1950) POSS I survey ($\delta \ga -33^\circ$), since our 
entire approach fails farther south.  USNO-A is a positional (and
photometric) catalog, which is constructed by matching blue and
red photographic plates.  In order to minimize the number of
artifacts masquerading as stars, USNO-A requires blue and red detections
with position differences less than $2''$.  Hence, in the non-POSS I areas
of the sky, where the blue and red plates were taken many years apart,
all stars with sufficiently high proper motion
will necessarily be absent.  For example, if the plates were
taken 8 years apart, then the catalog will not contain any
stars with $\mu>250\,\masyr$.  This does not pose any problem
for regions covered by POSS I because its blue and red plates
were taken sequentially.  Farther south, however, the plate
epochs differ by of order a decade.  There is a further wrinkle
here.  The first version of the USNO-A catalog, USNO-A1 (USNO-A1.0),
is constructed by scanning POSS I plates all the way to their
southern limit ($\delta\sim -33^\circ$) and supplements POSS I with 
additional southern surveys only south of this limit.  However, USNO-A2 (USNO-A2.0),
which has somewhat better
precision, exploits southern catalogs in place of POSS I for
the region $\delta\la -20^\circ$.  Hence, we
conduct our search using USNO-A2 wherever possible, but 
supplement this with USNO-A1 where necessary.

\subsection{The ``Rectangle''
\label{sec:rectangle}}

	We divide the search into two principle
stages, which we dub the ``rectangle'' and the ``circle''.
In the rectangle stage, we first identify all USNO-A stars (NLTT candidates)
that lie
within a $16''\times 8''$ rectangle\footnote{We consider genuine USNO stars
throughout: entries added from Tycho are ignored.}
centered on the position
predicted by using the reported NLTT proper motion to propagate 
the (epoch 1950) NLTT catalogued position to the epoch of 
the POSS plate underlying USNO-A.   We then predict the position of
the 2MASS counterpart of each such USNO-A star 
under the assumption that its proper motion is as given by NLTT.
We use a map of 2MASS coverage to determine if the star
should lie within the 47\% of the sky covered by the second 2MASS incremental
release and, if it should, what the exact 2MASS epoch is at that
position.  We then query the 2MASS data base for all stars lying
within $5''$ of this position.  The size of this error circle is
influenced primarily by the errors in NLTT proper motions, which
are expected from Paper I to be about $25\,\masyr$ for stars in the 
rectangle.  Since the
difference in epochs is typically about 45 years, this translates
to a $1.\hskip-2pt ''1$ error in predicted position in each
dimension, far larger than the $\sim 130\,$mas and $\sim 250\,$mas
position errors in 2MASS and USNO-A, respectively.  A $5''$ 
circle will therefore capture all but the extreme outliers.  Ideally,
only the USNO-A candidate actually corresponding to the NLTT star will
have a 2MASS counterpart because static stars will not match to anything
in 2MASS.  In general,
this query can have 3 possible outcomes, and we can therefore define 3 
classes of stars: non-matches, unique matches, and multiple matches.

\subsection{Unique Matches
\label{sec:unique}}

\begin{figure}
\epsscale{0.7}
\plotone{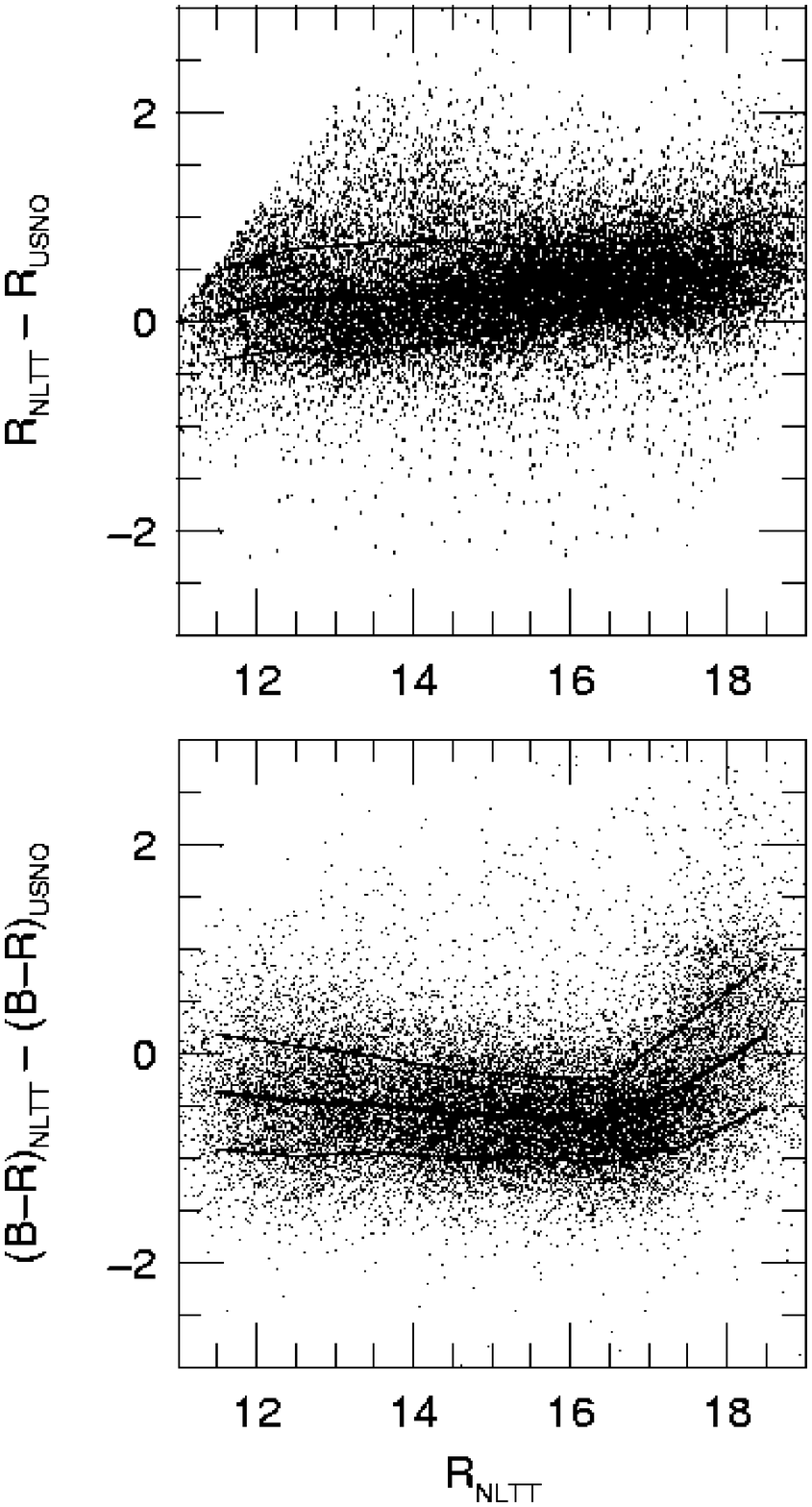}
\caption{\label{fig:brlu}
Differences between photographic photometry from NLTT and USNO for
stars in our catalog.  Upper panel shows the difference in
$R$ magnitude and bottom panel shows the difference in $(B-R)$ color.
The bold curves give the mean difference and the solid lines give
the $1\,\sigma$ scatter (with $3\,\sigma$ outliers removed from the fit)
in 1 mag bins.
This scatter ranges from 0.4 to 0.5 mag over most of the plot 
and is dominated by errors in NLTT.  Note that errors in $B$ and $R$
are highly correlated in both NLTT and USNO.
}\end{figure}

	Unique matches are those NLTT stars for which our
procedure generates one and only one 2MASS counterpart.
Because the area covered by the rectangle is quite small,
the overwhelming majority of unique matches should be genuine,
even though we have not yet imposed any magnitude constraint on
the selection procedure.  We can therefore use the unique matches
to study the color-color relations of counterparts in NLTT, USNO-A,
and 2MASS.  First, we find from the overlap of USNO-A1 and USNO-A2
stars that the $B_\usno$ mags are not on the same system in the two versions
of the catalog, which deviate by up to 1 mag.  We obtain
\begin{equation}
     B_{\rm A2} = B_{\rm A1} + 49.056 - 9.5613 B_{\rm A1} 
+0.669 B_{\rm A1}^2 -0.0198 B_{\rm A1}^3
+0.0002 B_{\rm A1}^4,\qquad (B_{\rm A1} > 13.07)
\label{eqn:rusno12}
\end{equation}
and $B_{\rm A2} = B_{\rm A1}$ otherwise.  Henceforth, when
we write $B_\usno$ it refers to magnitudes converted to the USNO-A2 system.
However, in the actual catalog we report the original USNO-A1 magnitudes
for stars identified in that catalog.  Next we compare USNO and NLTT
photometry.  Figure \ref{fig:brlu} shows this comparison (for our full
catalog, not just the rectangle matches) as a function of $R_\nltt$.  
Clearly, there are strong nonlinearities.  We model these by,
\begin{equation}
R_\usno = 0.9333 R_\nltt + 0.6932,\qquad (R_\nltt>10.39)
\label{eqn:rusnonltt}
\end{equation}
and $R_\usno = R_\nltt$ for $R_\nltt<10.39$, and by
$$(B-R)_\nltt - (B-R)_\usno = 0.2387 - 0.055 R_\nltt,\quad (R_\nltt <16.5)$$
\begin{equation}
(B-R)_\nltt - (B-R)_\usno = 7.3817 + 0.4067 R_\nltt,\quad (R_\nltt \geq16.5).
\label{eqn:brusnonltt}
\end{equation}
Note that $B_\nltt$ is called ``photographic magnitude'' in NLTT.

	Finally, we obtain very rough optical/IR color-color relations 
between USNO and 2MASS bands,
$$         R_\usno-J = -0.62 + 3.908(J-K_s), \quad
         B_\usno-J =  0.70 + 4.942(J-K_s), \qquad (J-K_s > 0.6), $$
\begin{equation}
         R_\usno-J =  0.37 + 1.536(J-K_s), \quad
         B_\usno-J =  0.73 + 3.305(J-K_s), \qquad (J-K_s < 0.6),
\label{eqn:usnojk}
\end{equation}
These last relations have a huge, 0.8 mag, scatter, and are not
even continuous.  They nonetheless are extremely useful in assessing
the quality of candidate matches.

\subsection{Multiple Matches
\label{sec:multiple}}

	Each multiple match is investigated and resolved by hand.
The primary cause of multiple matches is common proper motion (CPM) 
binaries.  Sometimes there are two UNSO-A stars that each match
to the same two 2MASS stars.  In such cases, there is always a
NLTT CPM binary.  This is because USNO-A does not resolve binaries
with separations $\la 5''$, whereas Luyten was able to resolve
these easily by eye.  Hence these cases are always resolved
unambiguously by making use of the NLTT supplementary notes on
CPM binaries.  More frequently, there are two 2MASS stars matched
to one USNO-A ``star'', which is actually an unresolved binary. 
In the great majority of these cases, NLTT shows a CPM binary,
so these are also easily resolved.  In such cases, however,
we cannot use the USNO-A position to obtain a reliable new
proper motion since it is a blend of two stars.  For such stars, 
our catalog has an improved 2000-epoch position
and new $J$ and $K_s$ photometry, but not an improved proper motion
nor better visual photometry (which is obtained from USNO).  
In some two dozen cases,
we resolve the multiple match by finding a probable CPM binary
that Luyten missed.  Some of these binary candidates can be confirmed
by investigating POSS I and POSS II images, but in most cases the
companion is too faint in the optical band or is blended in the POSS
images because the separation is of order $5''$ or less.
We will present these new binary companions in a future paper
on NLTT binaries.  Finally, for each resolved multiple match,
as well as for each unique match, we compare the $B_\usno$
and $R_\usno$ mags reported by USNO with the values that
would be predicted on the basis of NLTT and of 2MASS photometry
using the color-color relations established in 
\S\ \ref{sec:unique}.  Even though these relations have large
errors, they still help to identify wrong matches because
random pairs of unrelated stars are likely to differ by many
magnitudes, so by this standard even measurements with
0.8 mag errors can be very effective.
The outliers are regarded as suspicious and put aside, while 
the remainder are accepted.

\subsection{Rectangle Non-matches
\label{sec:nonmatch}}

	Operationally, there are two classes of non-matches: 
1) there are no USNO-A stars in the rectangle or 2) 
there are such stars but they all fail to match with 2MASS stars.  
Class (1) can occur because A) the NLTT position is in error
by more than the size of the rectangle (as we showed in Paper I
is the case for a sizeable minority of NLTT stars), or B) USNO failed
to detect the star. Class (2) can occur because A) the
NLTT position is in error and there just happens to be an
unrelated USNO star in the rectangle, B) the USNO star is
the correct match but 2MASS has failed to detect the star,
or C) the USNO star is correct but the NLTT proper motion
is so far off that the 2MASS counterpart lies outside the
$5''$ error circle.  We ultimately attempt to distinguish
among these five causes in \S\ \ref{sec:add_ids}
but for the moment assume that
that the correct USNO match lies outside the rectangle and
continue our search for it in the circle in \S\ \ref{sec:circle}.  

\subsection{The ``Circle''
\label{sec:circle}}

	We now remove from consideration all NLTT stars that
were successfully matched to USNO-A stars in the rectangle.
The great majority of the remainder are unmatched.  However, a 
small number were matched in the rectangle search, but are regarded
as questionable matches because they are photometric outliers as determined
using the relations in \S~\ref{sec:unique}.
We search for USNO-A/2MASS counterparts to both classes of stars 
by probing a circle of radius $2'$ in USNO-A centered on the NLTT
position (updated using the NLTT proper motion to the epoch of 
the USNO-A POSS plate).  Because the area being probed is
now 350 times larger than in the rectangle search, the probability
of false matches is proportionately larger.  We therefore take
several steps to reduce such false matches.

	First, we demand that 
$|R_\usno - R_\nltt|<2.2$.
These are approximately $5\,\sigma$
limits, as we established in \S\ \ref {sec:unique}.  (Some
NLTT stars do not contain an entry for $R_\nltt$, in which
case we us $B$ mags.)  Second, when we query the
2MASS data base at the 2MASS-epoch position, 
we ask whether the 2MASS star has any USNO-A
stars within $3''$ and if so what the separation and mags of the closest
one are.  If the 2MASS star is associated with a USNO-A star at
approximately the same position, we do not immediately eliminate 
the 2MASS star, but rather flag it.  If the association is 
real, the 2MASS star could not be the genuine
counterpart of the NLTT star because its proper motion would be
$\la 70\,\masyr$, far below the NLTT threshold.  However, the
association might be due to chance proximity of a random field
star.  At high latitude, the probability of such a chance
alignment is low, but in the difficult fields of the Galactic
plane, it is not.  We also do a reverse check: we ask whether there are
2MASS stars associated with USNO candidates (assuming no proper motion)
within $2''$.  The presence of a 2MASS star near the USNO 1950-epoch
positon will also indicate that the USNO candidate is not a NLTT star.

	For a large fraction of the difficult cases, we directly
consult the digitized sky survey (DSS).  This normally contains images
from two well-separated epochs (POSS I and II), and thus usually 
allows one to
spot the high proper-motion star.  However, for the southern
zones (approximately coincident with our USNO-A1 areas), the
two DSS epochs are very close in time, making it extremely difficult
to spot moving stars.  In some cases, we therefore consult the
USNO web site, which contains digitized POSS I, but the cycle time
for such searches is about 20 minutes, so this cannot be done in
every case.  This mostly affects regions very close to the Galactic
center.

	The multiple matches are more complex than in the
rectangle case.  We again have large numbers of CPM binaries.
These are usually easier to untangle than in the
rectangle case because they are typically farther apart and so are
resolved in USNO.  It is then only necessary to check that the
relative orientation of the 2MASS/USNO stars is the same
as that given by the NLTT notes.  However, there are now
a large number of spurious matches.  In a handful of cases,
these turn out to be CPM companions of NLTT stars that were
not found by Luyten.  We will present these in a forthcoming paper.  
Unlike the CPM binaries found in the rectangle, these are mostly well
enough separated that they are resolved in USNO.  The great
majority of the spurious matches are pairs of
unrelated USNO-A/2MASS stars.  We track these down using a
variety of photometric and astrometric indicators and by consulting
DSS, as discussed above.

	The nonmatches are of the same five classes that were described
in \S\ \ref{sec:nonmatch}.  These are sorted out in \S\ \ref{sec:add_ids}.

\subsection{Rectangle and Circle Summary
\label{sec:reccircsummary}}

	The entire procedure described above is carried out separately
for USNO-A2 ($\delta>-37.\hskip-2pt ^\circ 5$) and for 
USNO-A1 ($-37.\hskip-2pt ^\circ 5<\delta<-15^\circ$).  In each case,
we demand that the entry originate from POSS I.  The two catalogs are then 
combined, giving precedence to USNO-A2 when a match is found in both.
In practice, we find a fairly sharp boundary at $\delta\sim -20^\circ$.
Table \ref{tab:usno2mass} 
shows the number of matches in the four
categories (rectangles,circles)$\times$(unique,resolved), for each of
the two USNO catalogs, A1 and A2.  Of the 23,681 USNO/2MASS
matches, 18,442 (78\%) lie inside the rectangle.
The overwhelming majority of these are unique matches, and even the
3\% that had multiple matches were {\it relatively} easy to resolve.  Note
that even for the majority of cases in which the star lay in the circle,
there was only one USNO/2MASS pair whose separation was consistent with 
the NLTT proper motion.  Thus, the high quality of the NLTT PPMs, 
combined with the paucity of stars over most of the sky, allowed us
to match over 20,000 stars in two catalogs with very different passbands, 
at widely separated epochs, with ``only'' about a 1000 problem cases.

	However, these cases, which were mostly concentrated in Galactic plane,
were often quite difficult to resolve.  As is well-known, optical and 
infrared images of the plane are each extremely crowded, but often not with 
the same objects, and this fact generally gave rise to dozens, sometimes
more than 100, false matches for a single NLTT entry.  
The correct match had to be laboriously
identified from among these.  Thus, our method, which works amazingly well
away from the plane, tends to become extremely bogged down within it.
An alternative approach to finding proper motion stars in the plane, 
albeit one that would miss many red stars, would be to compare blue plates
in which the majority of background stars would be removed.  In 
\ref{sec:homosinb} we will see that that is exactly what Luyten did!

\section{{Comparison with PPM}
\label{sec:PPM}}

	Before trying to deal with the residual nonmatches from the
rectangle/circle search, we first reconcile our results with the search
for matches with PPM catalogs (bright stars), which was described in Paper I.
Disagreements will yield clues to possible problems with each 
of the two approaches as well as an estimate of the false identification
rate for cases in which the NLTT star is identified in
only one of the two methods, which constitute the vast majority.  

	We begin by examining the 27 cases for which a USNO star was found
in the rectangle with no counterpart being found in 2MASS, but for which
the NLTT star was identified in a PPM catalog.  In 23 of these 27, the
USNO identification was correct.  For 18, the 2MASS match failed because
the NLTT proper motion error was larger than $100\,\masyr$, and therefore put
the 2MASS counterpart outside the $5''$ error circle.  For the remaining five,
the star was not in 2MASS despite the fact that it was in an area nominally
covered by the 2MASS release.  In the remaining 4 cases, the USNO 
identification was incorrect.  In three of these, the USNO/NLTT magnitude
differences were too large to make a plausible match (but recall that the
reason for permitting such large differences is the prospective confirmation
from a 2MASS counterpart, which in this case is absent).  That is, among
the 24 cases for which the NLTT/USNO rectangle match seemed plausible, in 23 
cases it was correct.  The fact that there are few false USNO/NLTT
matches in the rectangle, even when there is no identifiable 2MASS counterpart,
is the basis for the aggressive effort to identify nonmatches
that we describe in \S\ \ref{sec:add_ids}.

	Next we examine the 1922 cases for which the NLTT star was identified
in both the PPM and USNO/2MASS searches.  In 53 of these, the two methods
identified different stars.  However, 28 of these 53 involved problems 
with the mis-identification of binary components, all of which would have
come to light and been corrected when we examine CPM binaries in \S\
\ref{sec:cpm}, even if they had not been in this small overlap sample.
Hence, we focus on the remaining 25 cases.

	Ten of these 25 were caused by problems in PPM.  For one, a bad
Starnet proper motion for the correct optical star led to the 
identification of the wrong 2MASS counterpart.  In all nine others,
the true optical star was too faint to appear in the PPM catalogs, and
another match was found as the search criteria were gradually degraded.
This is exactly the problem that was foreseen for this method.  For
the areas of the sky that are covered by our faint search, this problem was 
``automatically'' corrected by this comparison.  However, since our
faint search has been carried out only in the overlap of the 2MASS release 
with POSS I, which covers 44\% of the sky, we expect that there remain about
$10\times(100 - 44)/44 \sim 13$ such wrong identifications over the rest
of the sky.  Eventually, when all of 2MASS is released, the those with
$\delta>-33^\circ$ will be spotted.  However, this will leave the 
remaining 23\% of the
sky unchecked in which, we extrapolate, there are only 5 false identifications.

	The other 15 cases were caused by problems with the USNO/2MASS
search.  In 6 cases, a USNO star was found inside the rectangle and
a 2MASS counterpart was found to this star, but there was actually
a much better match outside the rectangle.  This is the major potential
difficulty with our method: if we find what looks like a reasonable
match in the rectangle, we do not look farther.  Otherwise our search
would become unwieldy.  For four of these six, the USNO/2MASS magnitude
discrepancy was large.  Based on this finding, we 
eventually removed all matches for which this
discrepancy exceeds 3 mag.  In another 6 cases, USNO did not contain the 
correct star and so the search found another star and managed to match
it with a 2MASS star.  However, all 6 cases are eliminated by our
subsequently adopted cut on USNO/2MASS magnitude discrepancies.  In
two cases, NLTT position or proper-motion transcription errors led
to the identification of a spurious USNO/2MASS pair, and in one case
we chose the wrong one of two neighboring 2MASS stars as the counterpart
of the correctly identified USNO star.

	That is, after implementing our cut on USNO/2MASS magnitude
discrepancies, there were only 5 out of 1922 cases, where we obtained
a false identification.

	Finally, there are 80 stars that were found in Tycho-2 or
Starnet but not in the USNO/2MASS search.  Of these, 15 are unresolved
binaries.  Another 34 result from severe (usually transcription) 
errors in NLTT: either they lie outside the $2'$ USNO search circle (16)
or they had very large proper-motion errors (18).  Another 7 are so
bright that they fail to appear in USNO, and 9 others are in the
dense Galactic center fields where we have an especially difficult time
identifying stars.  Three others had various miscellaneous causes and the
remaining 12 had no identifiable cause.  The binaries would be caught in 
the investigation summarized in \S\ \ref{sec:cpm}.  In \S\ \ref{sec:add_ids},
we describe our method to recover most of the stars with large proper motion
errors.  Bright stars are not a problem beyond the magnitude range of
the PPM surveys.  The remaining 40 cases constitute somewhat over
2\% of the 1800 Tycho-2+Starnet matches with USNO/2MASS.  

	In brief, based on this comparison of  the results of the PPM search
and the USNO/2MASS search, we believe that our false identification rate
is well below 1\% and we anticipate that we will fail to match about 2--3\%
of NLTT stars (until we reach the flux levels where USNO and 2MASS 
become seriously incomplete).

\section{{Additional Matches}
\label{sec:add_ids}}

	Based on our analysis of the cases in which the PPM search
found the NLTT star but the USNO/2MASS search did not, we conduct
four additional types of searches, which we dub ``annulus'', ``2MASS-only'',
``USNO-only'', and ``CPM''.  The first is aimed at finding 2MASS counterparts
to USNO stars in cases for which the NLTT proper motion is seriously in error,
including the possibility that there is a transcription error in the 
direction of proper motion.  Hence the search is conducted in an annulus
around the position of the unmatched USNO star.  

	The second and third
searches are designed for cases in which the star is, for some reason,
absent from USNO or 2MASS.  A major cause of missed stars in these
catalogs is simply that they are too faint.  Since NLTT contains a
substantial number of stars to $V\sim 19$ and 2MASS is complete roughly
to $J\sim 16$, some blue stars with $V-J<3$ will be lost.  Fortunately,
while there are a substantial number of stars with $V-J<3$ in NLTT,
the vast majority of them are intrinsically bright main-sequence stars
or subdwarfs and so, given the proper motion limit
$\mu > 180\,\masyr = 85\,\kms/(100\,\pc)$, are not affected by the 2MASS
magnitude limit.  However, some subdwarfs and many white dwarfs are
affected.

	USNO requires that a star be identified on both the blue and
red plates, and so misses faint stars of extreme color, primarily faint
red stars, which are found mainly at the bottom of the main sequence.
However, it should be noted that both USNO and 2MASS occassionally miss
much brighter stars, primarily because of saturation by or confusion
with a neighboring star, but sometimes for no reason that we can identify.
That is, occassionally we find an isolated, reasonably bright star
on the images that is simply not in the catalog.  Although the fraction
of such cases is small, this small fraction is multiplied by 23,000 in
our search.

	The last search is for stars that NLTT identifies as CPM companions
to stars that we have identified in either the USNO/2MASS or PPM search.
As we will show in \S\ \ref{sec:binerrs}, 
NLTT {\it relative} astrometry of these
pairs is usually very accurate, even in cases for which its absolute
astrometry is poor.  Hence these companions are usually easy to find.
This search technique can therefore provide an independent measurement of the
false identification rate in our overall USNO/2MASS search.  This is very
important because it is possible that the low (0.3\%) misidentification
rate that we found in \S\ \ref{sec:PPM} could have been a result of the
fact that we were testing only the brightest of the USNO/2MASS stars.

	The most important step we take to avoid contamination in the
first three searches is to essentially restrict them to the rectangle
where, as discussed in \S\ \ref{sec:PPM}, contamination is very low.
Stars whose NLTT position is outside the rectangle will therefore not
be identified in these searches.

\subsection{{The Annulus}
\label{sec:annulus}}

	In the PPM search described in Paper I, we found many examples 
of transcription errors of all types in NLTT, including in the position 
angle of the proper motion.  We therefore search for
2MASS stars in an annulus around the position of all USNO counterparts 
lying within the NLTT rectangle but without previous 2MASS matches.
To also allow for a substantial error in the NLTT value for the amplitude 
of the proper motion, $\mu_\nltt$,
we set the inner and outer and outer radii of the
annulus at $\theta_\pm = (T_{\rm 2MASS} - T_\usno)(\mu_\nltt \pm 100\,\masyr)$.
Here $T_{\rm 2MASS}$ and $T_\usno$ are the respective epochs of the 2MASS
and USNO observations.  Queries of the 2MASS database allow one to ask
for the nearest USNO star within $5''$.  We accept 2MASS stars only
if they do not have a plausible USNO counterpart as determined from the
consistency of the optical/IR colors of the two catalog objects relative
to the predictions of equation (\ref{eqn:usnojk}) and their relative
position offset.  We also check that the
USNO rectangle star does not have a plausible 2MASS counterpart at
small separation, which would also indicate that it is not a proper
motion star.  Occassionally we find a 2MASS counterpart to the USNO
star near the outer edge of the $5''$ search radius which, 
given the $\sim 45$ year baseline, 
could only be a real counterpart 
if the proper motion were $\mu\la 100\,\masyr$.
In this case we permit the identification provided that the position
angle is reasonably consistent with NLTT.  In total, we find 50 stars
in the annulus, or 0.2\%, substantially smaller than the 1\% (18/1800)
that were identified in the overlap with Tycho-2+Starnet in \S~\ref{sec:PPM}.
This may be because the PPM search was more aggressive than the one we have
conducted here (since it could make use of additional proper-motion
information).

\subsection{{2MASS-Only Identifications}
\label{sec:2mass_only}}

	When no USNO/2MASS match is found either in the circle or the
rectangle, and when there is no USNO candidate within the rectangle,
we hypothesize that the NLTT position is correct to within
the precision of the rectangle, but that the star is absent from USNO.
We therefore predict the position of the star at the 2MASS epoch
assuming that the NLTT 1950 position and proper motion are correct.
We then search for 2MASS counterparts of the NLTT star in a radius of $12''$ 
around this position to allow for NLTT PPM errors.  As in the annulus
search, we reject 2MASS stars that have plausible USNO counterparts.
Of course, the result of this search yields only a position and IR photometry,
but not a new proper motion.  In all, this search yields 862 2MASS-only
identifications.  

\subsection{{USNO-Only Identifications}
\label{sec:usno_only}}

	When a valid USNO counterpart (one without an associated 2MASS
star having consistent optical/IR colors and lying within a few arcsec)
of the NLTT star is found within the rectangle, but it is not matched
with a 2MASS star either in the original search or the annulus search,
we assume that the star is absent from 2MASS and accept the 
USNO/NLTT identification, provided that the USNO/NLTT magnitudes are
consistent.  In these cases, we obtain a new position and
new optical photometry, but no new proper motion.  In all, this search
yields 409 USNO-only identifications.  We argue in \S\ \ref{sec:sghighlat}
that most of these stars are WDs.

\subsection{{Common Proper Motion Companions}
\label{sec:cpm}}

	The notes to NLTT contain references to more than 2400
pairs that Luyten considered to be CPM binaries or possible binaries.
Unfortunately, it would require substantial advances in artificial
intelligence software to extract all of this information in a completely
automated way.  The biggest single problem is that, for some reason,
Luyten refused to name a large numbers of the objects in his catalog.
In these cases, he will have a note to one star such as ``comp.\ to prec.\
star''.  Sometimes the companion will be the preceeding star, but very
often it will be several preceeding, or occassionally following.  Often,
when there is a reference to a named star, the meaning of the reference
is clear only from context.  In brief, it seemed impossible to make
sense of these references in an automated way.  Mercifully, for stellar
pairs that  Luyten
believed to be CPM binaries, he almost always recorded the
same proper motion for both components, even when (as we will show in 
\S\ \ref{sec:binreal}) 
he was able to measure the difference.  This permitted us to
adopt a different approach.

	First, we create our own naming system: each NLTT star is
named for its sequential number in the elctronic version of the
catalog, from 1 to 58,845.  Next,
we run a program to find all notes that might plausibly be interpreted
as indicating that the star is a component of a binary or multiple system.
Normally, these contain a separation and position angle relative to a 
primary, but sometimes they are more ambiguous.
We then find all NLTT stars within one degree of the noted star 
that have identical
NLTT proper motions (magnitude and position angle). If any of these stars
had been identified in either the PPM or USNO/2MASS searches, we print out the
catalog entries as well as all the notes that are associated with all of 
these stars.  If two or more of these stars have been identified, we
also print out the separations and position angles of all pairs.  We
then review this output, tracking down discrepancies when we have
two or more identifications and our relative astrometry is in conflict
with NLTT's, and we interactively query the 2MASS data base to search
for additional matches when the companion is not already in our data base.
After updating our catalog with these results, we run a second program that 
finds only those notes that it can parse well enough to determine a
predicted separation and position angle.  It assumes that this vector
separation applies to all pairs of NLTT stars with the same proper motion
and lying within 1 degree.  If our catalog contains entry pairs that
disagree with this separation by more than 20\% or with the position
angle by more than $20^\circ$, or if we do not have an entry for one
component but do have an entry with 2MASS identification for the other,
it flags this system.  All such flagged systems are reviewed and the
2MASS data base is searched for unmatched companions.  

	This search
yields 137 CPM companion identifications, of which 59 are corrections
to previously incorrect 2MASS identifications.  Of these 59, 47 are
components of binaries for which our USNO/2MASS search found the
same star for both components.  Our algorithm is prone to such errors
because it treats each component separately, and we intended
to sort out these discrepancies at the time we reviewed all binary matches.
They therefore do not represent real mismatches.  However, the remaining
12 are genuine misidentifications: if these stars, which were treated
by our algorithm as individuals, did not happen to belong to CPM binaries,
we would never have recognized the error.  Since, there are a total of
1235 stars in CPM binaries that were matched to USNO/2MASS, this represents
a false-identification rate of about 1\%.  Alternatively, one might want
to consider only those stars in CPM binaries with separations greater than
$10''$ in order to probe a sample that is not severely affected by
confusion with a nearby high-proper motion source.  This would be a fairer
proxy to the conditions that the algorithm faced for the majority of single
stars.  In this case, we get 6 misidentifications out of 784 stars.
Or, for a more conservative $30''$, we get 4 misidentifications out
of 442 stars.  That is, our estimated misidentification rate is
consistently 1\%.  This rate is about three times higher than the 
0.3\% rate we estimated from comparison with PPM stars.  Most likely,
this is because brighter stars are rarer, and hence the possibility
that a random unrelated USNO/2MASS pair will mimic a given high-proper
star is smaller.  In any event, we consider a 1\% misidentification
rate to be quite good.

	Since the CPM search was conducted after the search for 2MASS-only
stars, we can use it to check the reliability of the 2MASS-only 
identifications.  There are 34 2MASS-only stars that lie in CPM binaries
{\it in addition} to the 137 companion identifications discussed above.
The relative positions of all 34 stars are consistent with the values
in NLTT, and in this sense all 34 ``pass the test''.  However,
of these 34, 26 are from 13 CPM pairs both components of which are identified
only in 2MASS.  Since we processed both stars from a pair simultaneously,
we had knowledge of their CPM nature and offsets, and so these 26 do not
constitute the a fair test.  The remaining 8 stars are companions of
independently identified NLTT stars, and so are a good proxy for single
stars.  The statistical significance of this test is modest because of
small number statistics, but at least it has the right sign.  One of the
8 is a companion to a USNO-only star, which does not have an independent
proper motion.  The remaining 7 are additional CPM companions each of
whose proper motions can be inferred from that of
its companion.  This brings the total of such
CPM companions to 144.

\section{{Description of Catalog}
\label{sec:catalog}}

	The catalog contains information grouped in six sections,
1) summary, 2) NLTT, 3) source identifications, 4) USNO, 5) 2MASS,
6) binaries, plus the last column which is an internal reference for
debugging.

\subsection{{Summary Information}
\label{sec:suminfo}}

This section contains 11 entries: 1) the NLTT number (drawn consequentively
from 1 to 58,845), 2) a letter code `A', `B', or `C' if the NLTT ``star''
has been resolved into several sources, 3) $\alpha$ (2000, epoch and equinox), 
4) $\delta$ (2000),
5) $\mu_\alpha$, 6) $\mu_\delta$, 7) $\sigma(\mu_\alpha)$, 8) 
$\sigma(\mu_\delta)$ (all four in arcsec yr$^{-1}$), 9) $V$, 10) $V-J$,
11) 3-digit source code.

The three digits of the source code refer to the sources of the position,
proper motion, and $V$ photometry.  1 = Hipparcos, 2 = Tycho-2, 
3 = Tycho Double Star Catalog (TDSC, \citealt{tdsc}), 4 = Starnet,
5 = USNO/2MASS, 6 = NLTT, 7 = USNO (for position) or common proper motion 
companion (for proper motion).  More specifically, ``555'' means
2MASS based position, USNO based $V$ photometry, and USNO/2MASS based
proper motion.

	The (2000) position has been evolved forward from whatever epoch
it was measured using the adopted proper motion.  When the positon source
is Hipparcos, Tycho-2, or TDSC, the position is given in degrees 
to 6 digits, otherwise to 5 digits.  
For proper motions derived from a PPM catalog, the errors are adopted
from that catalog.  Proper motions from USNO/2MASS determinations 
have an estimated error of $5.5\,\masyr$ is accordance with the results from 
\S\ \ref{sec:poserr}, 
below.  NLTT proper motions are $20\,\masyr$ as found in \S\ 
\ref{sec:pmerrwide}.
CPM binary companions (without other astrometry) are not assigned an error,
and zeros are entered into the error fields.  

As described in Paper I, $V$ refers to the Johnson $V$ entry for Hipparcos,
Tycho $V$ for Tycho-2 and TDSC, and Guide-star catalog $R$ for Starnet.
The conversion from USNO photometry is given by \citet{nearbylens},
\begin{equation}
\label{eqn:usnov}
      V = R_\usno + 0.23 + 0.32(B-R)_\usno.
\end{equation}
We remind the reader that for USNO-A1 photometry, we first convert to
USNO-A2 using equation (\ref{eqn:rusno12}) before applying equation
(\ref{eqn:usnov}).  When NLTT photometry is used, $V$ is evaluated using
equations (\ref{eqn:rusnonltt}) and (\ref{eqn:brusnonltt}) to convert
to USNO mags, and then applying equation (\ref{eqn:usnov}).  In the 
rare cases for which NLTT photometry is employed and one of the two
bands is not reported, a color of $(B-R)_\nltt = 1$ is assumed.
No effort has been made to ``de-combine'' photometry in the case of unresolved
binaries.  For example, if a binary is resolved into two stars in
2MASS, but is unresolved in USNO, then different $J$ band measurements
will be reported for the two stars, but both with have the same,
combined-light $V$ photometry.  The $V-J$ color reported in field 10
will be the simple difference of these two numbers.  Similarly, if
the NLTT ``star'' is resolved by TDSC but not 2MASS, then the
$V$ light will be partitioned between the two stars but not the $J$
light.  Finally, note that if no $J$ band photometry is available
(whether because of saturation, faintness, or the star being in an area
outside the second incremental 2MASS release) $V-J$ is given as $-9.$

When multiple sources of information are available, the priority for what
is presented in the summary is as follows. 
Positions: 3,1,2,5,4,7,6;  
Proper Motions: 3,2,1,5,4,7,6; 
Photometry: 3,1,2,5,4,6.

As discussed in Paper I, Tycho-2 proper motions are given precedence over
Hipparcos primarily because they better reflect the long-term motion
when the stars are affected by internal binary motions, but also because
at faint magnitudes they are generally
more precise.  When Hipparcos proper motions are given, it is often because
the star is so faint that it does not show up in Tycho.  In this case,
the nominal Hipparcos errors are often quite large and true errors can
be even larger.  We found a handful of cases by chance in which the Hipparcos
proper motion was grossly in error and we removed the Hipparcos entry
and substituted the USNO/2MASS value.  However, we made no systematic
effort to identify bad Hipparcos proper motions.

Only stars for which we are providing additional information are recorded
in the catalog.  There are 36,020 entries for 35,662 NLTT stars including
a total of 723 entries for 361 NLTT ``stars'' that have been resolved in TDSC.

\subsection{{NLTT Information}
\label{sec:nlttinfo}}

The next 6 columns give information taken from NLTT, namely
12) $\alpha$ (2000), 13) $\delta$ (2000), 14) $\mu_\alpha$, 15) $\mu_\delta$,
16) $B_\nltt$, 17) $R_\nltt$.  The coordinates and proper motions are
precessed from the original 1950 equinox to 2000, and the position is updated
to 2000 epoch using the NLTT proper motion.

\subsection{{Source Information}
\label{sec:sourceinfo}}

The next 2 columns give source information.  Column 18 is the Hipparcos number
(0 if not in Hipparcos).  Column 19 is the identifier from TDSC, Tycho-2,
or Starnet, whichever was used to determine the position in columns 3 and 4.
When the position comes from Hipparcos, 2MASS, or USNO, or NLTT, ``null'' is
entered in this field with one exception: when a Starnet measurement
has been superseded 
by a 2MASS measurement, the Starnet identifier has been retained for ease
of recovery of this source.  It can easily be determined that the summary
information comes from 2MASS because the first digit in field 11 
will be a ``5''.

\subsection{{USNO Information}
\label{sec:usnoinfo}}

The next six fields give USNO information:
20) Integer RA, 21) Integer DEC, 22) $B_{\rm A1\ or\ A2}$, 
23) $R_{\rm A1\ or\ A2}$, 
24) USNO Epoch,
25) 3-digit search-history code.  The Integer RA and DEC together serve
as a unique USNO identifier since that is the form RA and DEC are given in
the original USNO-A1 and USNO-A2 releases.  They can also be converted into 
degree $\alpha$ and $\delta$ (at the USNO epoch) using the formulae: 
$\alpha$ = (Integer RA)/360000, $\delta$ = (Integer DEC)$/360000 - 90$. 
Regarding the 3-digit search history code, the first
digit tells which USNO catalog the entry is from: 1 = USNO-A1, 2 = USNO-A2.
The second tells whether the USNO source was found in the rectangle (1) or
the circle (2).  The third tells whether it was a unique match (1), or had
to be resolved by hand from among several possible matches (2).
If there is no USNO information, all of these fields are set to zero.

\subsection{{2MASS Information}
\label{sec:2massinfo}}

The next six fields contain 2MASS information:
26) $\alpha$, 27) $\delta$ (both at 2MASS Epoch), 28) $J$, 29) $H$,
30) $K_s$, 31) 2MASS Epoch.  If no 2MASS data are available, all fields
are replaced by zeros.  If there are 2MASS data, but not for a particular
magnitude measurement, that value is replaced by $-9$.

\subsection{{Binary Information}
\label{sec:binaryinfo}}

The next six fields contain information about binarity:
31) binarity indicator, 32) NLTT number of binary companion, 33) NLTT
estimated separation, 34) NLTT estimated position angle, 35) our
estimated separation, 36) our estimated position angle.  Regarding
the binarity indicator, 0 means NLTT does not regard this as a binary.
Otherwise, it is a NLTT binary and the 
indicator is set according to whether the companion
is (2) or is not (1) in our catalog.  The NLTT estimates of the separation
position angle come from the NLTT Notes.  Our estimates come from the
difference of the 2000 positions of the two stars.  In cases for which
the companion is not in our catalog, the fields with ``our'' separation
and position angle are replaced by values found from the difference of
the NLTT coordinates (i.e., fields 12 and 13).  As discussed in \S\ 
\ref{sec:cpm}, the companion numbers are based on what we think is
obviously what Luyten intended, rather than what was literally written
down.  However, no effort has been made to clean up any other transcription
errors, even when these are equally obvious.  No binary information is
recorded in these fields about NLTT ``stars'' that were resolved by TDSC.
Rather, the reader should recognize each of those binaries from the upper case
letter appended to its NLTT number (column 2).

The final column (37) is an integer that is used in the program that
assembles the catalog from the various subcatalogs that are described in
Paper I and in \S\ \ref{sec:strategy} and \S\ \ref{sec:add_ids} of this
paper.  It is useful mainly to us, but for completeness:
1 = PPM+2MASS, 2 = USNO/2MASS, 3 = CPM, 4 (not used), 5 = 2MASS-only,
6 = PPM (no 2MASS), 7 = Annulus match, 8 = USNO-only.

The Fortran format statement for the catalog record is:\hfil\break\noindent
(i5,a1,2f11.6,2f8.4,2f7.4,2f6.2,1x,3i1,2f10.5,2f8.4,2f5.1,i7,1x,a12,
\hfil\break\noindent
2i10,2f5.1,f9.3,i2,2i1,2f10.5,3f7.3,f9.3,i2,i6,f7.1,f6.1,f7.1,f6.1,i2)

\section{{Reduced Proper Motion Diagram}
\label{sec:rpm}}

\begin{figure}
\plotone{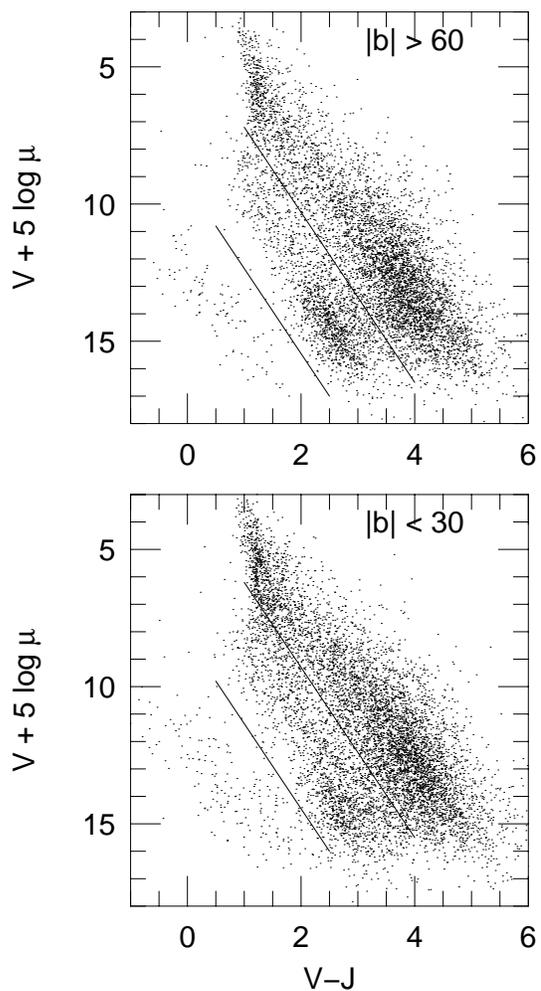}
\caption{\label{fig:rpm}
Reduced proper motion (RPM) diagrams for all stars in our catalog having
$J$ photometry, restricted respectively to Galactic polar and equatorial
regions.  In each diagram, main-sequence stars, subdwarfs, and white
dwarfs are well separated.  The line segments are drawn by eye to
delineate the separations.  The lines are parallel but offset by 1 magnitude
in the two diagrams because the mean speed of stars is substantially higher
toward the poles.  Eqs.\ (\ref{eqn:eta}) and (\ref{eqn:discrim}) extrapolate
from the equations of these lines to give delineators at any Galactic latitude.
}\end{figure}

	Perhaps the single most important product of a proper motion
survey is the reduced proper motion (RPM) diagram.  In addition to its
general utility for separating out different populations of stars,
the RPM diagram will prove extremely useful for characterizing the
properties both of the NLTT catalog and of the catalog we are presenting
here.
In Figure \ref{fig:rpm}, we plot the $V$ band RPM, 
\begin{equation}
V_\rpm = V + 5 \log\mu,
\label{eqn:vrpm}
\end{equation}
against $V-J$ color for respectively high-latitude $(|b|>60^\circ)$ and
low-latitude $(|b|<30^\circ)$ stars in our catalog.  In each region of the
sky, the subdwarfs (SDs) are clearly separated from 
the main sequence (MS) and the white dwarfs (WDs), but the divisions are
at slightly different places because stars seen toward the poles tend to
be moving faster.  We make a linear interpolation between the
two diagrams (noting that the first has $\langle |\sin b|\rangle = 0.93$
and the second, $\langle |\sin b|\rangle = 0.25$) and arrive at a single
discriminator $\eta$, by which we classify stars as a function of their
position in the RPM diagram and their Galactic latitude,
\begin{equation}
\eta(V_\rpm,V-J,\sin b) =  V_\rpm - 3.1(V-J) - 1.47 |\sin b| - 2.73.
\label{eqn:eta}
\end{equation}
The classification is:
\begin{equation}
{\rm MS:}\quad \eta<0,\qquad {\rm SD:}\quad 0<\eta<5.15,\qquad {\rm WD:}\quad 
\eta>5.15\ .
\label{eqn:discrim}
\end{equation}
Notice, for example, that in the high-latitude diagram 
$\eta(V_\rpm,V-J,0.93)=0$ is the equation of the upper separator line, while
$\eta(V_\rpm,V-J,0.93)=5.15$ is the equation of the lower one.  We
will use equations (\ref{eqn:eta}) and (\ref{eqn:discrim}) to extract
subsamples from different stellar populations throughout the rest of this 
paper.  As we have elsewhere shown \citep{rpm}, it was not possible to do
this with the original NLTT RPM diagram. 

\begin{figure}
\epsscale{1.0}
\plotone{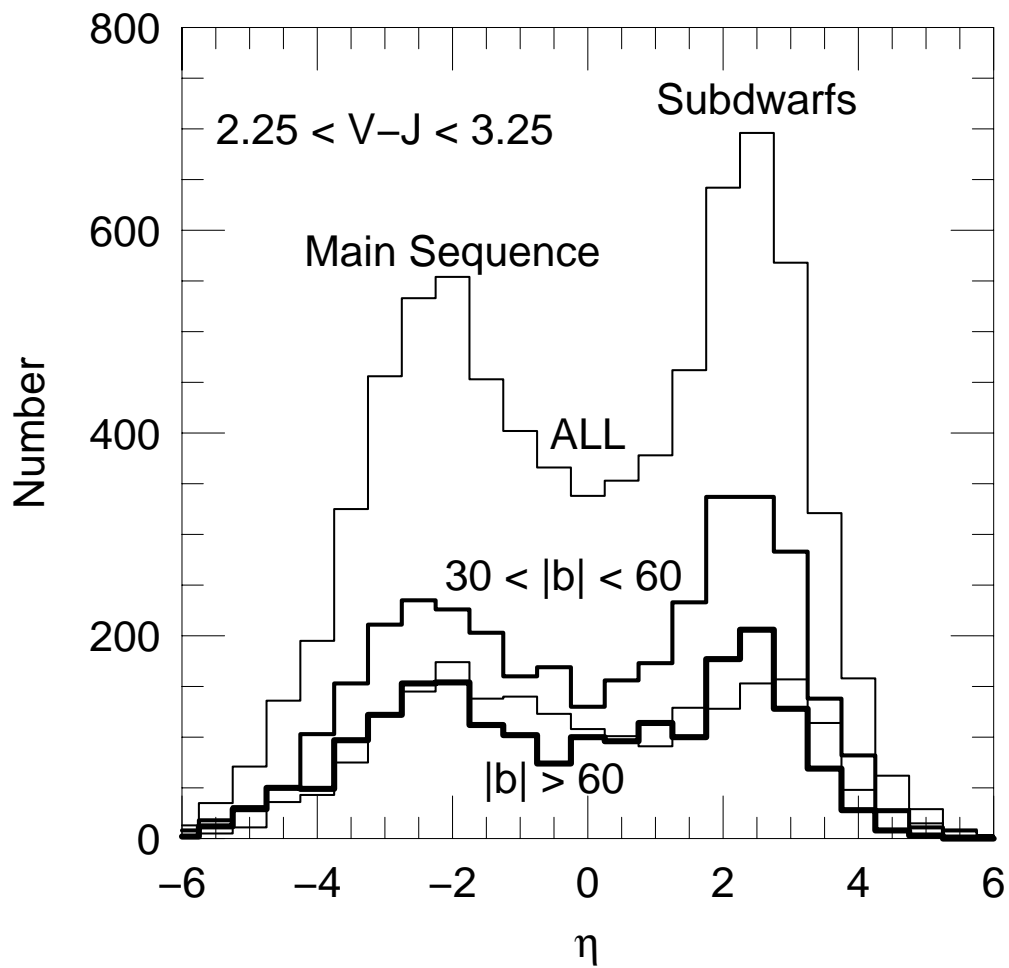}
\caption{\label{fig:eta}
Distribution of $\eta$, (eq.\ [\ref{eqn:eta}]),
which discriminates among main-sequence stars, subdwarfs, and white dwarfs, 
broken down into 3 latitude bins, and for all stars in our catalog,
subject to the restriction $2.25<V-J<3.25$.  In this color range there
are essentially no white dwarfs.  The FWHM for the subdwarfs is about 2 mag,
and for the main sequence about 3 mag.  The peaks are separated by about
5 mag.
}\end{figure}

	Figure \ref{fig:eta} shows the distribution of $\eta$ in
the color range $2.25<V-J<3.25$ for 3 ranges of Galactic latitude and
for all combined.  The diagram shows clear peaks for the MS and the SDs
with FWHM of about 3 magnitudes and 2 magnitudes respectively.  An
important question that we will address in \S\ \ref{sec:charSG} is: to
what extent is this width intrinsic to $\eta$ and to what extent is
it due to measurement error?  The figure shows that the separation
between the MS and the SDs is not perfect.  If one wanted a very clean
sample of SDs, one would choose $\eta>1$ (or perhaps slightly higher),
rather than $\eta>0$.  Or alternatively, one could use a more sophisticated
approach that made use of both components of the proper motion and both
Galactic coordinates.  However, for purposes of studying the properies
of the catalog, this level of cleanness is not required, and we will
simply divide up the sample according to the conventions of equation
(\ref{eqn:discrim}).

\section{{Characteristics of NLTT}
\label{sec:charNLTT}}

	Here we use our improved measurements of NLTT stars to characterize
the precision of its positions, proper motions, photometry, and 
binary separations, as well as the reality of its CPM identifications.
We defer issues of completeness to \S~\ref{sec:complete}.

\subsection{{Positions}
\label{sec:poserr}}

\begin{figure}
\epsscale{0.75}
\plotone{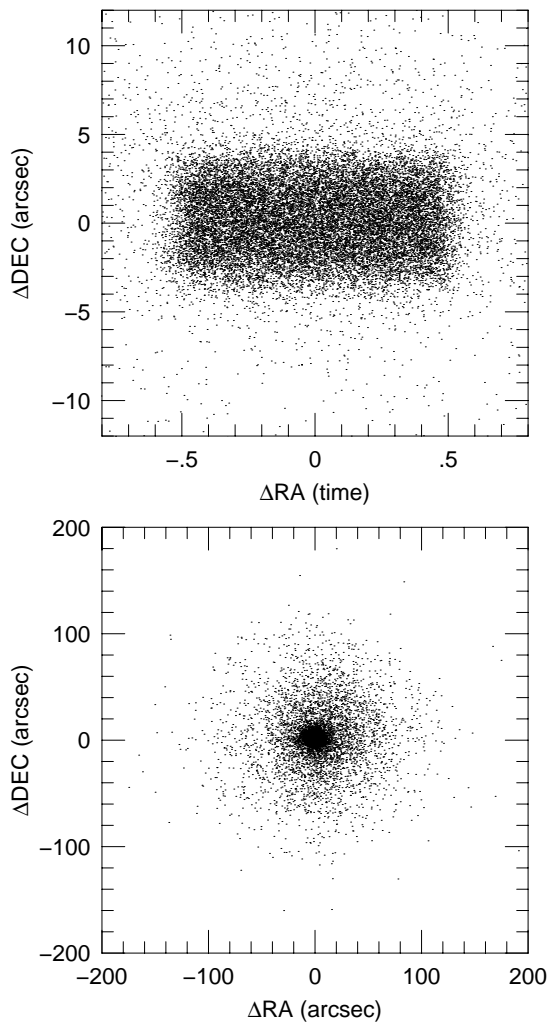}
\caption{\label{fig:pos_above_11}
Differences between NLTT listed position and true position (as determined
by propagating the star back to its 1950 position using the 2000 position
and proper motion from our catalog) for all faint, $V>11$, stars.  The
rectangle in the upper panel arises because, for most of his catalog,
Luyten actually measured positions to about $1''$ but only recorded them 
to 1 s of time and $6''$ of arc.  However, the lower panel shows that there is
a substantial halo of outliers at least out to our search radius of $2'$.
Compare to Figs.\ 1 and 2 from \citet{paper1}.
}\end{figure}

	Figure \ref{fig:pos_above_11} shows the differences between
the 1950 epoch positions of NLTT and those predicted based on the 2000
positions and proper motions in our catalog.  It is restricted to faint
($V>11$) stars to maximize the potential contrast with Figures 1 and 2
from Paper I, which show the same quantities for bright NLTT stars
found in Tycho-2.  In fact, there is no qualitative difference between
the bright and faint position errors.  The majority of the residuals
are roughly uniformly distributed in a rectangle whose dimensions are
set by the decimal truncation of the NLTT entries.  The ``fuzziness''
of the rectangle edges gives an estimate of the underlying measurement
errors.  As was true of the Paper I bright sample, 
this is about $1''$.  Figure \ref{fig:pos_above_11}b
shows that the heavily populated rectangle is surrounded by a diffuse halo
extending out to at least $2'$, which was the limit of our faint search
in most cases.  Nevertheless, taken as a whole, the positions are extremely
good.  These good positions played a critical role in our ability to
identify the vast majority of NLTT stars in USNO and 2MASS in an automatic
fashion.  Note that the claim by \citet{bakos} that the typical 
LHS (subset of NLTT) position
errors are larger than $10''$ is manifestly not true.

\begin{figure}
\epsscale{1.0}
\plotone{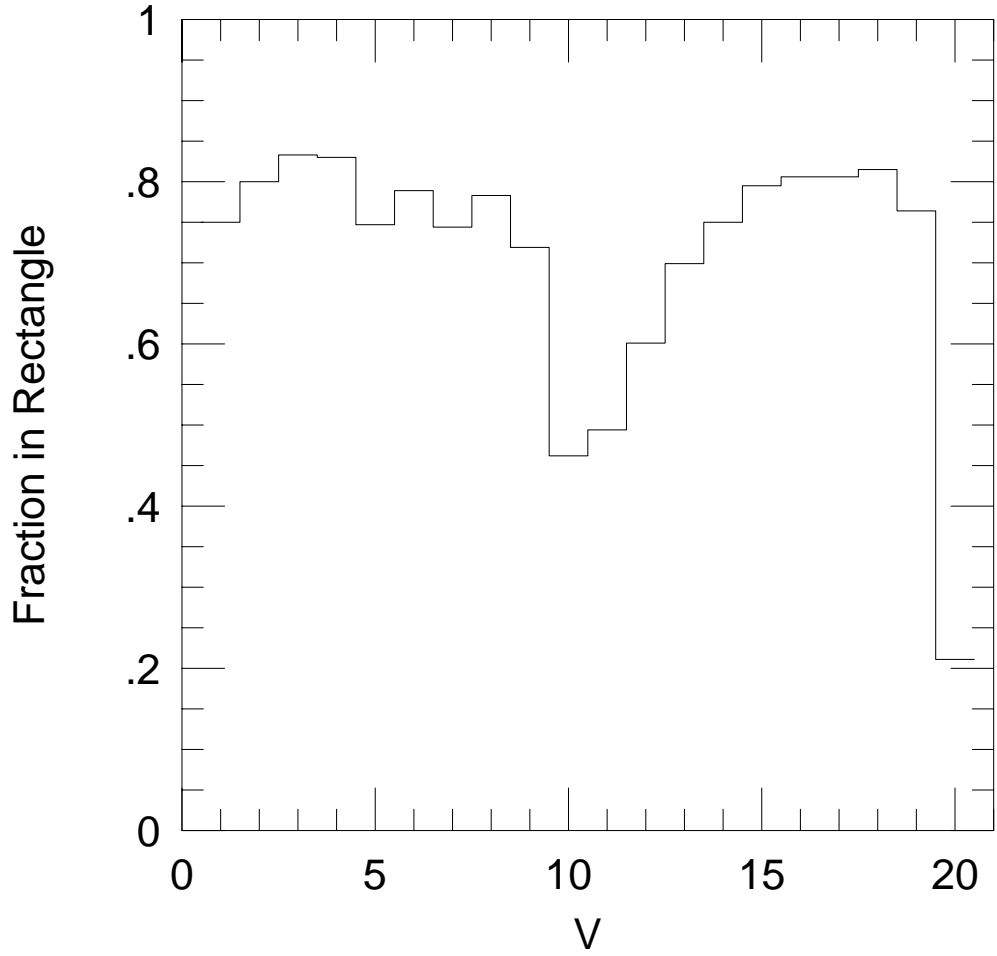}
\caption{\label{fig:rectmag}
Fraction of NLTT stars found in a rectangle centered on the NLTT position and
with dimensions $(15''\cos\delta + 2'')\times(6'' + 2'')$ as a function
of $V$ magnitude.  The size is set to allow for $1''$ errors in addition
to error caused by decimal truncation of the catalog entries.  The feature
at $V\sim 10$ is probably caused by difficulty doing astrometry for 
stars saturated on plates.
}\end{figure}

	We further investigate the magnitude dependence of the NLTT position
errors in Figure \ref{fig:rectmag}, which shows the fraction of NLTT
stars whose actual position lies within a rectangle of dimensions
$(15''\cos\delta + 2'')\times (6'' + 2'')$ and centered on the NLTT recorded
position.  The first term in each dimension is the imprecision created
by decimal truncation and the second allows for $1''$ measurement errors at
each edge of the rectangle.
The fraction remains mostly constant at about 80\% over 19 magnitudes, except
for a pronounced feature suddenly dipping to 45\% at $V=10$ and then
gradually climbing back to 80\% at $V=15$.  We speculate that the
brighter stars $V\leq 9$ were already well studied and that Luyten made
use of literature data on these.  For the remainder of the stars he made
his own measurements from plates.  Saturation badly affects such measurements
at $V=10$, but gradually ameliorates at fainter mags.

\begin{figure}
\plotone{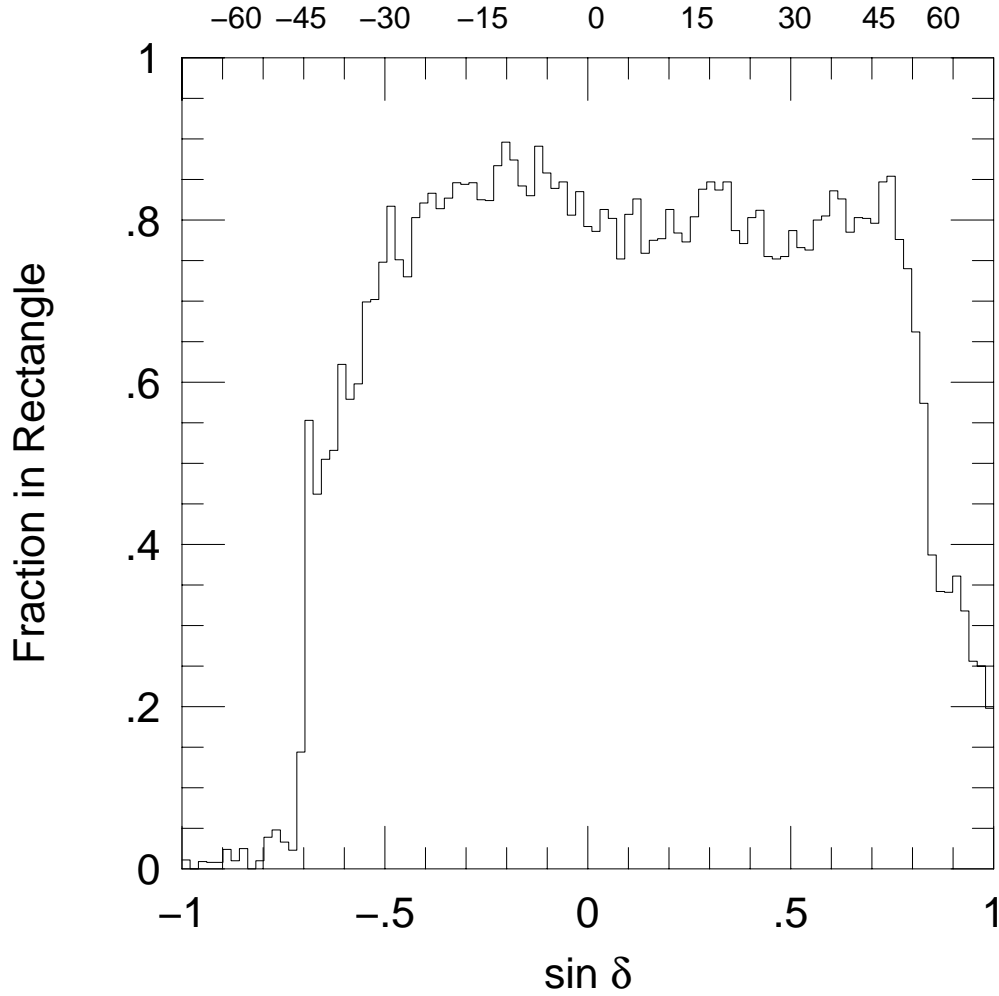}
\caption{\label{fig:rectlat}
Fraction of NLTT stars found in a rectangle centered on the NLTT position and
with dimensions $(15''\cos\delta + 2'')\times(6'' + 2'')$ as a function of
of declination.  The fraction is very high over most of our catalog
area, $\delta > -33^\circ$, but deteriorates drastically to the south.
}\end{figure}

	Figure \ref{fig:rectlat} shows the rectangle fraction as a function
of declination.  This figure has a number of important features.  Over most
of the sky, $-30^\circ < \delta < 50^\circ$, the fraction is roughly 80\%.
As expected, it drops to essentially zero for $\delta < -45^\circ$ where
NLTT records positions only to 6 seconds of time and $1'$ of arc.  Less 
expected, however, are the rapid dropoffs for $\delta < -30^\circ$ and 
$\delta>50^\circ$.  The former dropoff has significant negative consequences
for the prospects of extending our catalog to the south, $\delta < -33^\circ$,
as we discuss in \S\ \ref{sec:south}.

\subsection{{Proper motions}
\label{sec:pmerr}}

\begin{figure}
\plotone{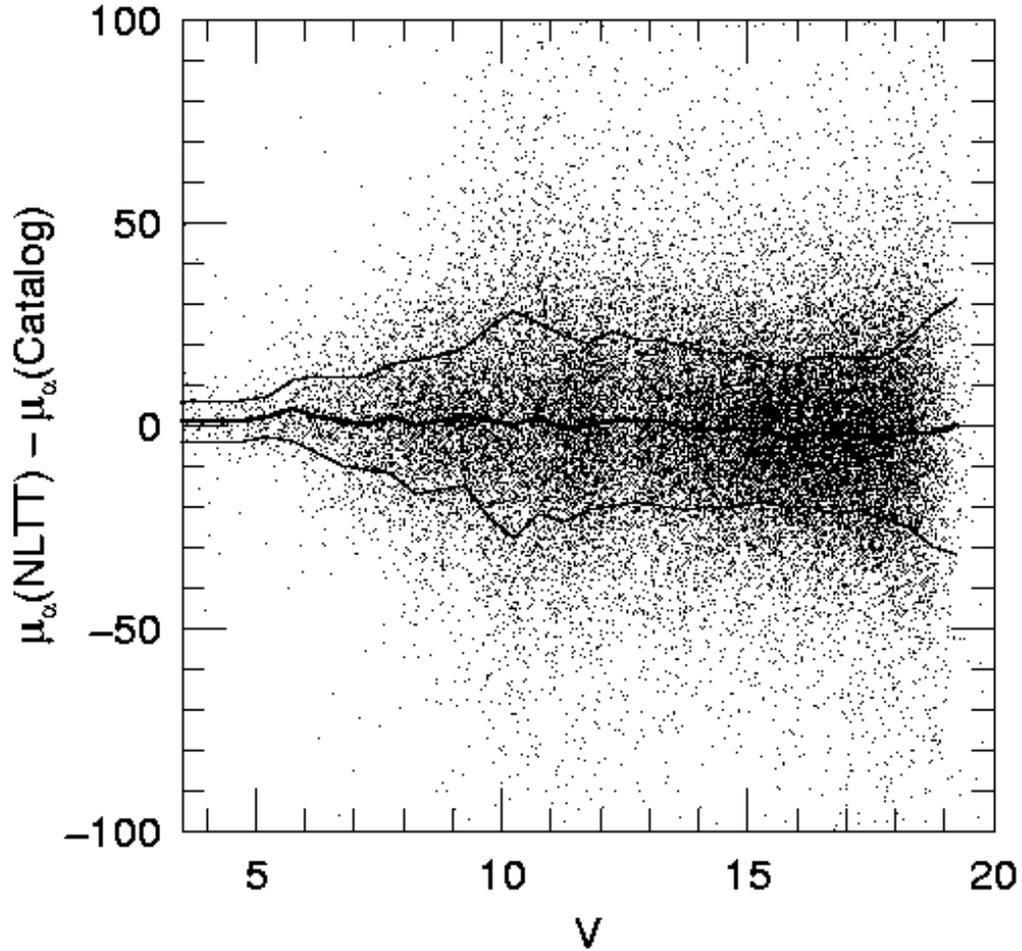}
\caption{\label{fig:pmx}
Differences between proper motions in the $\alpha$ direction as given
by NLTT and our catalog.  The bold line shows the mean difference,
which is consistent with 0.  The solid lines indicate the $1\,\sigma$
scatter (with $3\sigma$ outliers removed from the fit) in 0.5 mag bins.  
The NLTT
errors are typically $20\,\masyr$ in the range $11<V<18$. The pronounced
scatter at $V\sim 10$ may be related to the feature seen at the same
place in Fig.\ \ref{fig:rectmag}.
}\end{figure}

	Figure \ref{fig:pmx} shows the differences between proper motions
in the $\alpha$ direction as given by NLTT and our values.  The diagram
for the $\delta$ direction looks extremely similar and is not shown.
Binaries with separations closer than $10''$
are excluded from this plot because our measurements, at least, can be
corrupted due to blending.  
The bracketing lines show the $1\,\sigma$ scatter (with
$3\,\sigma$ outliers excluded).  At bright magnitudes, NLTT errors 
are extremely small, then rise typically to $20\,\masyr$.  However, there
is a curious ``bump'' of higher errors at $V\sim 10$, which is reminiscent of
the degradation in positions (see \S\ \ref{sec:poserr}) and which may have the
same cause.  At the very faintest mags, $V\ga 18$, the errors deteriorate
toward $30\,\masyr$.  

	The fraction of $3\,\sigma$ outliers excluded from the fit is fairly 
constant at $\sim 3\%$, which is much higher that would be characteristic 
Gaussian noise.  These outliers may be due to transcription errors in NLTT, 
misidentifications of NLTT stars by us, or possibly other causes.  Whatever
the cause, the reader should not apply Gaussian norms to the interpretation
of the tails of this error distribution.

\subsection{{Photometry}
\label{sec:photerr}}

	Figure \ref{fig:brlu} shows the differences between NLTT and
USNO photographic magnitudes and colors as a function of magitude.
The $1\,\sigma$ scatter is typically 0.4-0.5 mag in both diagrams,
except that the color scatter deteriorates to 0.7 mag at the faint end.
This scatter, which is mostly contributed by NLTT, partly accounts for
the poor performance of the NLTT reduced proper motion (RPM) diagram
in discriminating among different stellar populations 
However, as we have shown \citep{rpm}, even the RPM diagram constructed
with superior USNO photometry does not distinguish main-sequence stars
from subdwarfs.  The fundamental problem is the shortness of the 
photographic $(B-R)$ color baseline.

\subsection{{Binary Relative Astrometry}
\label{sec:binerrs}}

\begin{figure}
\plotone{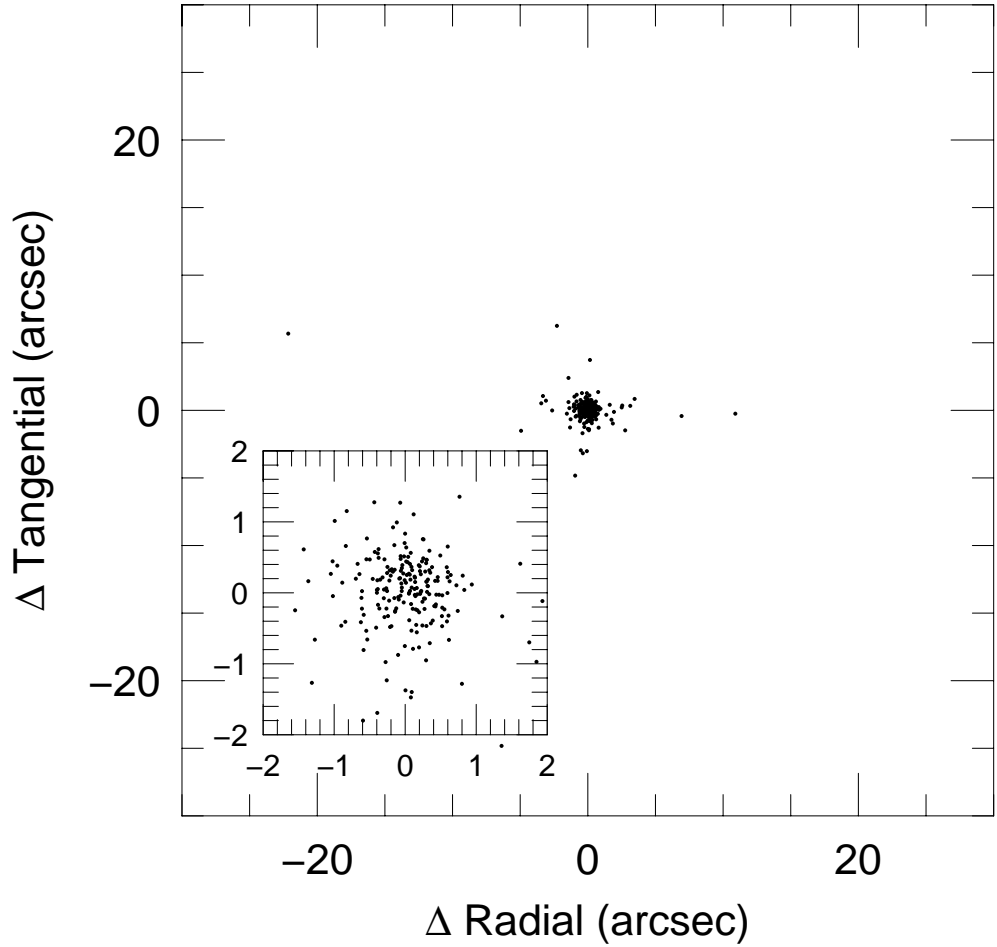}
\caption{\label{fig:cpm_pos}
Difference between separation vectors of 339 binaries with separations
$10''<\Delta \theta < 57''$ as given by NLTT and our catalog.  The 
$x$-coordinate
is along the direction of the separation and the $y$-coordinate 
is perpendicular
to it.  The great majority of NLTT separation vectors are accurate to $1''$.
See inset.
}\end{figure}

	Figure \ref{fig:cpm_pos} shows the difference in 339 separation vectors
between binary components as given by NLTT versus our astrometry.  The
sample is restricted to separations $\Delta\theta>10''$ to avoid problems
in our astrometry due to confusion, and $\Delta\theta<57''$ to avoid the
regime in which NLTT errors induced by truncation of the position angle
(given to integer degree) become larger than the errors induced by the
truncation of the separation (given to integer arcsec).  The inset shows
that this relative astrometry is usually very good, better than $1''$.
A number of the large outliers are due to transciption errors in NLTT.
This highly precise relative astrometry permitted us to reliably identify 
NLTT CPM components even when USNO data were missing.

\subsection{{Reality of NLTT CPM Binaries}
\label{sec:binreal}}

	Our improved proper motions permit us to better determine the
reality of the CPM binaries listed in the NLTT notes.  Figure \ref{fig:cpm_pm}
shows the differences in component proper motions for 468 NLTT binaries
with separations $\Delta \theta > 10''$ (again, to avoid problems with our
astrometry due to confusion).  

	Of course, one does not expect the components of CPM binaries to have 
exactly the same proper motion.  First they have orbital motion, which
for face-on circular orbits induces a relative motion 
$\Delta\mu = 2\pi(M/D^3\Delta\theta)^{1/2}$, where $D$ is the distance, $M$ is
the total mass, $\Delta\theta$ is the angular separation, with units of
pc, $M_\odot$, arcsec, and years.  Even if they had the same physical velocity,
their proper motions would differ because the components are at different
distances and because we see different components of motion projected on
the sky.  The order of these differences combined is 
$\Delta\mu/\mu\sim \Delta\theta$, in radians.
We add these orbital and projection effects in quadrature and show them
as error bars in Figure \ref{fig:cpm_pm}.  We estimate distances using the
brighter (but non-WD) component.  We classify it as a SD or MS star
according to equation (\ref{eqn:discrim}), and then assign it an
absolute magnitude $M_V = 2.7(V-J) + 2.1$ or $M_V = 2.09(V-J) + 2.33$
in the respective cases.  In addition, there is a
$6\,\masyr$ error (see \S\ \ref{sec:pmerrnarrow})
in our relative proper-motion measurements that must be
added in quadrature to the errors shown in the figure.  We have not done
this to avoid overwhelming the intrinsic scatter.  Also, we have not placed
any error bars on the points lying below $12\,\masyr$ 
(i.e., $2\,\sigma$) to avoid
clutter and because in our view these CPM identifications can be accepted 
with good confidence.

	The figure shows that the overwhelming majority of NLTT CPM binaries
with separations $\Delta\theta <50''$ are real, but that close to half of
those with $\Delta\theta>100''$ are unrelated optical pairs.  In fact,
a number of these with proper motion differences $\Delta\mu>100\,\masyr$
could have been excluded by Luyten at the $2.5\,\sigma$ level given his
$20\,\masyr$ precision in each direction and for each component.  However,
he evidently decided to err on the side of not missing potential CPM
binaries.

\section{{Characteristics of New Catalog}
\label{sec:charSG}}

	The new data that we provide for NLTT stars dramatically reduces
their astrometric and photometric errors.  Here we characterize these
errors.

\subsection{{Proper Motions Errors: Wide Angle}
\label{sec:pmerrwide}}

	Once bright stars are identified with PPM catalog stars, they
acquire the proper motions, and with them the proper motion errors,
given in those catalogs.  These vary but are generally of order a few
$\masyr$.  Faint-star proper motions are obtained by taking the difference
of 2MASS and USNO positions.  Again, these vary in quality but typically
have errors of order 130 and 250 mas respectively.  Given the $\sim 45$ year
baseline, we expect proper motion errors of order $6\,\masyr$.  Nevertheless,
one would like an independent experimental confirmation of this estimate.

	To obtain this, we compare in Figure \ref{fig:tycu2mpm} our USNO/2MASS
proper motions with Tycho-2 proper motions for 1179 stars for which we
have measurements from both.  We exclude for this purpose binaries closer
than $10''$ because, as we have emphasized several times, they can
be corrupted by blending.  The scatter, which is dominated by USNO/2MASS
errors, is high at bright magnitudes and then plateaus at $V\sim 10$ at
about $6\,\masyr$ in each component.  The poorer quality for bright
stars is due to the problems that USNO astrometry has in dealing with
saturated stars.  Because we do not use USNO/2MASS proper motions for
bright stars, our primary interest is in the asymptotic behavior of
the error envelope toward faint magntitudes.  Taking an average over the
bins $V>10$ and taking account of the small contribution to the scatter
due to Tycho-2 errors, we find average errors of 
\begin{equation}
\sigma_\mu = 5.5\,\masyr, \qquad \rm (wide\ angle),
\label{eqn:widangpm}
\end{equation}
in each direction.  These error bars are calculated excluding $3\,\sigma$
outliers, which constitute almost 5\% of the points.  Thus, as in the case
of NLTT, these proper-motion errors have strong non-Gaussian tails.

\subsection{{Proper Motions Errors: Narrow Angle}
\label{sec:pmerrnarrow}}

	USNO astrometry, which dominates the proper-motion error budget,
is more accurate on small scales than large scales.  On scales of several
degrees it suffers from errors in the plate solutions while on smaller scales
it is limited by centroiding errors.  For some applications, notably
studying the reality of binaries or their internal motions
(see \S\ \ref{sec:binreal}), it is the small scale errors that are 
relevant.  

\begin{figure}
\plotone{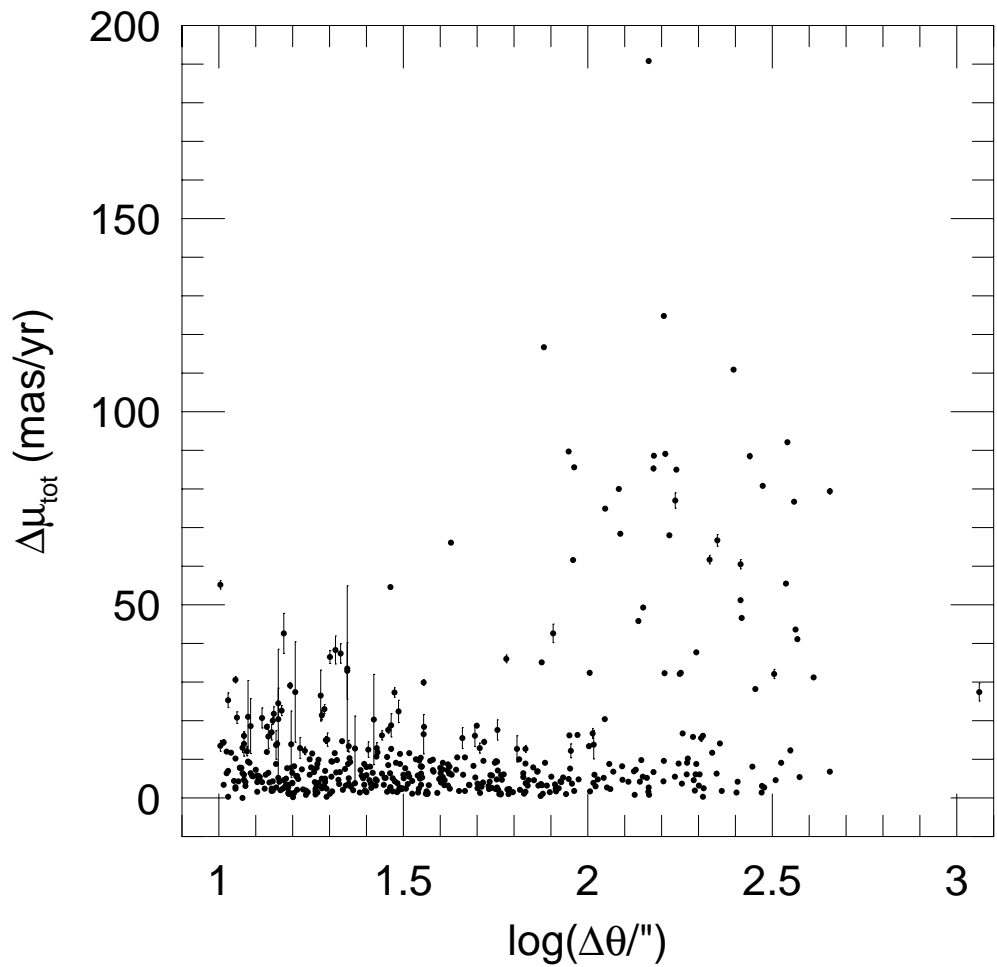}
\caption{\label{fig:cpm_pm}
Magnitude of the difference in the relative vector proper motion of
binary members as given by NLTT and our catalog.  Error bars reflect
the effects of internal binary motion at small separations, $\Delta\theta$,
and projection effects at large separations.  They do not include our
measurement error of $6\,\masyr$.  Points with $\Delta\mu <12\,\masyr$
are likely to be genuine pairs and do not have error bars to avoid clutter.
At large separations, many NLTT ``CPM'' binaries are actually optical pairs.
}\end{figure}

\begin{figure}
\epsscale{0.75}
\plotone{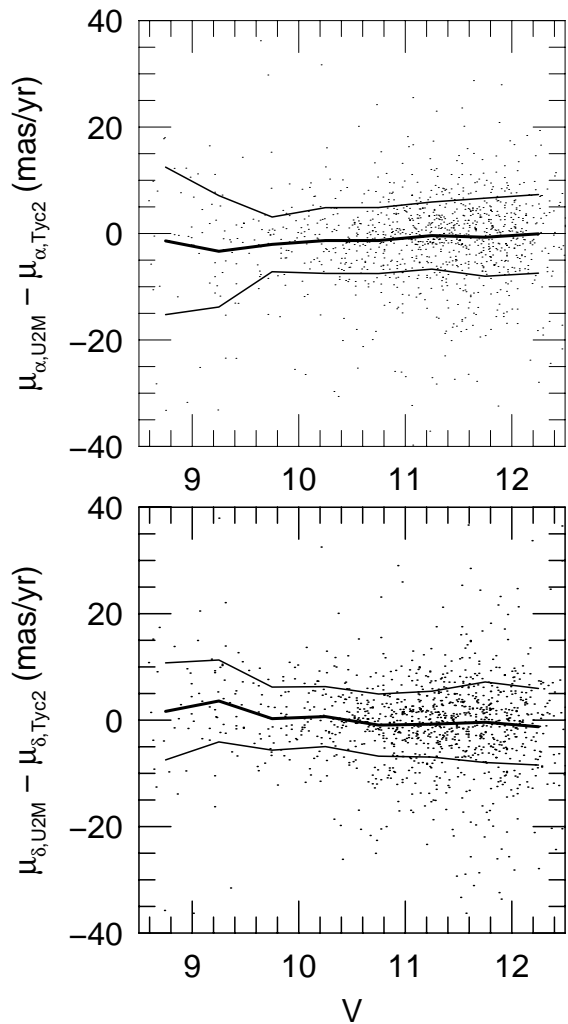}
\caption{\label{fig:tycu2mpm}
Differences between proper motions as given by Tycho-2 and our measurements
based on identifying the stars in USNO and 2MASS.  Most of the difference
is due to errors in the USNO/2MASS measurement, which are thereby evaluated
to be $5.5\,\masyr$ for $V>10$.  
The bold curve shows the mean difference (consistent
with 0) and the solid curves show the $1\,\sigma$ scatter (with $3\,\sigma$
outliers omitted from the fit) in 0.5 mag bins.  
Errors are more severe at bright magnitudes
because saturation adversely affects USNO astrometry.
}\end{figure}

\begin{figure}
\epsscale{1.0}
\plotone{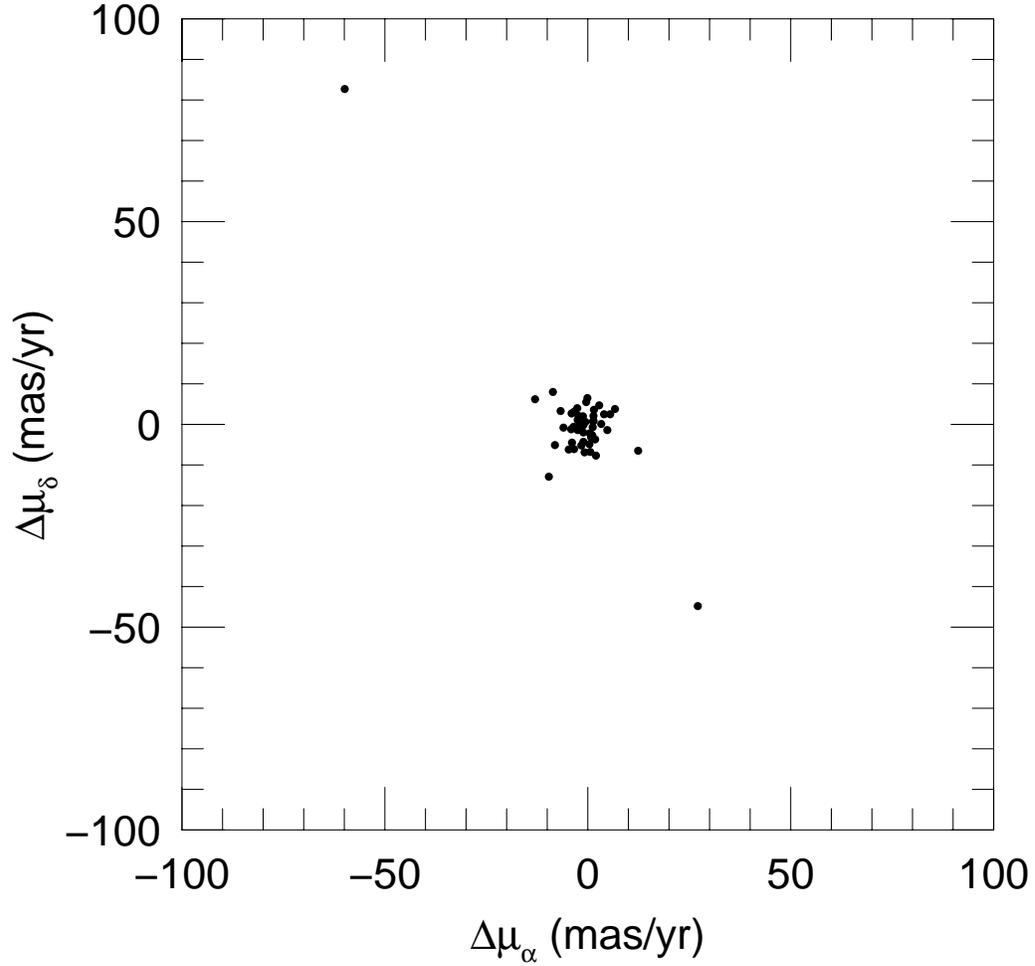}
\caption{\label{fig:cpm_halo}
Difference in the vector relative proper motion of the components of 
52 subdwarf binaries with separations $\Delta \theta>10''$.  For physical
pairs the real relative proper motion is very close to 0, so these
differences provide an estimate of our narrow-angle proper-motion
errors, $3\,\masyr$ (with the two outliers omitted).
}\end{figure}

	To determine these narrow-angle errors,
we plot in Figure \ref{fig:cpm_halo} the
difference in the proper motions of the two components of 52 SD
binaries (both components have $0<\eta<5.15$, see eqs.\ 
[\ref{eqn:eta}] and [\ref{eqn:discrim}]) with separations 
$\Delta\theta>10''$.  We choose
SDs because they are more distant, which in turn implies that these
binaries will have wide physical separations (and so be relatively unaffected
by internal motions) without being widely separated on the sky (and
so prone to contamination by optical binaries -- see Fig.\ \ref{fig:cpm_pm}). 
In a proper-motion selected sample, the mean distance is proportional to the 
mean transverse speed, which is of order five times larger for SDs than
MS stars.  
The figure shows a tight clustering of points with two outliers at
$\ga 10\,\sigma$, which are either not physical binaries or have extremely
bad proper-motion measurements.  After excluding these, we obtain a scatter
in the two directions of $4.4\,\masyr$  and $4.2\,\masyr$.  Since each 
results from the combination of two proper motions measurements, we derive
a narrow-angle proper-motion error of
\begin{equation}
\sigma_\mu = 3.0\,\masyr, \qquad \rm (narrow\ angle).
\label{eqn:narangpm}
\end{equation}
Hence, the rms error in the magnitude of the proper motion difference of
a binary is $6\,\masyr$, which was the value we adopted in analyzing
Figure \ref{fig:cpm_pm}.

\begin{figure}
\plotone{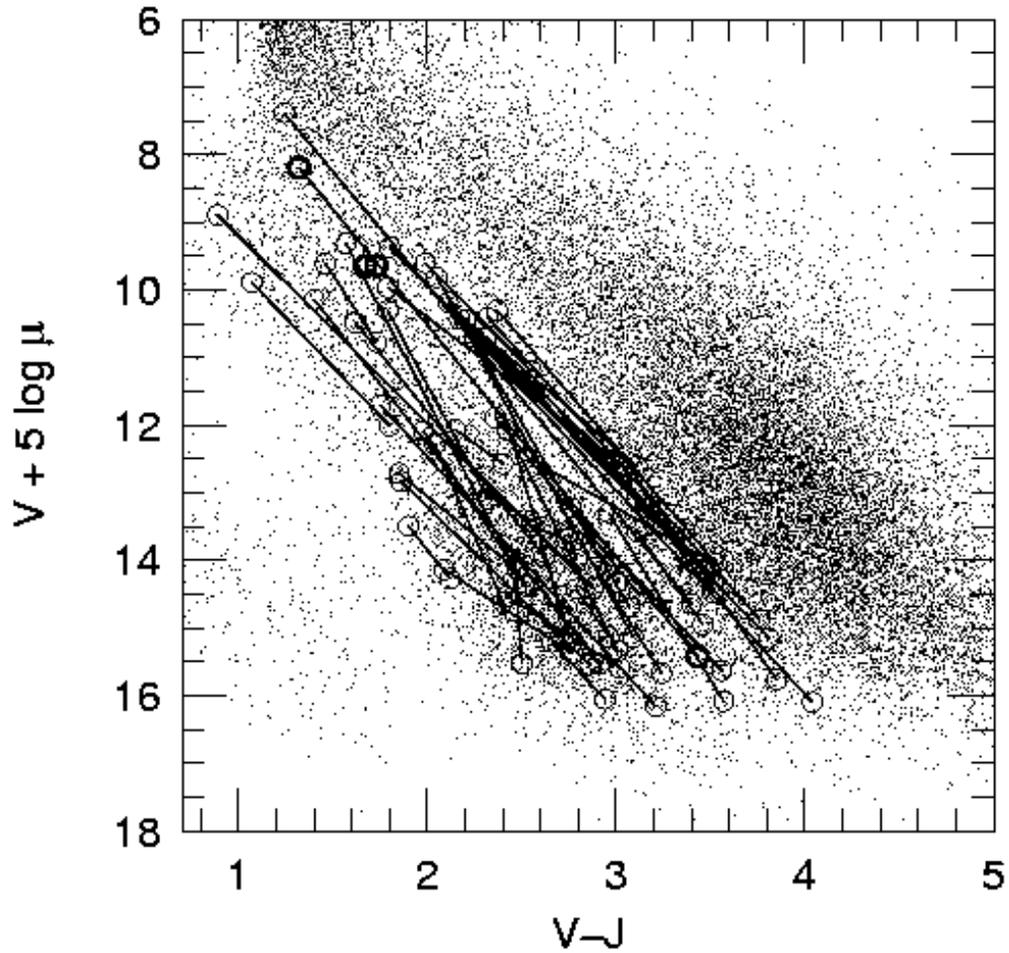}
\caption{\label{fig:cpm_rpm}
Reduced proper motion (RPM) diagram of all stars in our catalog with the
components of the 52 binaries from Fig.\ \ref{fig:cpm_halo} shown
as large circles connected by line segments.  One expects that physical
pairs will lie at roughly the same distance below the main sequence,
which the great majority of these do.  Notice that the two outlier binaries
from Fig.\ \ref{fig:cpm_halo} (shown in bold circles) also have this 
property, and so may be real pairs.
}\end{figure}

	Figure \ref{fig:cpm_rpm} shows these 52 binaries on a RPM diagram, 
together with solid lines connecting the two components.  Note that
in the great majority of cases, the two components lie about equal
distances below the MS, as would be expected for two stars of similar
chemical composition and transverse velocity but different luminosities.  
This also applies to
the two outliers from Figure \ref{fig:cpm_pm} (shown as bold circles),
so they may in fact be genuine pairs with very large proper-motion errors.

\subsection{{Photometry Errors and the RPM Diagram}
\label{sec:rpm_err}}

	To understand how measurement errors affect the RPM diagram
(Fig.\ \ref{fig:rpm}) we focus on the discriminant $\eta$.  The form of
$\eta$ given in equation (\ref{eqn:eta}) is not conducive to this analysis
because $V$ appears in both axes, the color $V-J$ and the RPM $V_\rpm$.
Hence, we first rewrite this equation as
\begin{equation}
\eta =  3.1 J - 2.1 V + 5\log\mu - 1.47 |\sin b| - 2.73.
\label{eqn:rpmalt}
\end{equation}

	The standard against which we must measure the errors in $\eta$ is the
width of the narrower (SD) peak, which is about 2 mag FWHM, corresponding
to about 0.85 mag Gaussian half-width.  At the proper-motion limit of NLTT
($180\,\masyr$), the $5.5\,\masyr$ proper motion errors reported in
\S\ \ref{sec:pmerrwide}, induce an error in $\eta$ of 0.07 mag, which is
completely negligible on this scale.  Indeed, even the NLTT proper-motion
errors, which are more than 3 times larger, would not contribute significantly.

	For the overwhelming majority of our stars, the $J$ band photometry
errors are 0.02--0.03 mag.  Even at 0.1 mag (which applies to only about 1\% of
the stars), the errors would not be significant relative to width of the
SD distribution.

	Because bright stars are saturated in 2MASS, the great majority
of stars with RPM measurements derive their $V$ data from USNO's photographic
photometry.  For these, the errors are approximately 0.25 which, from 
equation (\ref{eqn:rpmalt}) induces an error in $\eta$ of 0.53 mag.
This does provide a significant, albeit nondominant contributiion to
the width.  If it could be removed, for example by obtaining CCD photometry
for all NLTT stars, the width of the SD $\eta$ distribution would be reduced
by about 20\%.

\subsection{{Position Errors}
\label{sec:pos_err}}

	For some applications, notably the search for nearby-star microlensing
candidates \citep{nearbylens}, the position errors are the most important.
For the fainter stars $V>11$, our positions come from 2MASS and have a
typical accuracy of $130\,$mas as of about 1998.  These will deteriorate
(in quadrature) at a rate of about $5.5\,\masyr$ due to accumalated proper
motion error.

\section{{Completeness}
\label{sec:complete}}

	We have deferred the problem of assessing the completeness of
both NLTT and our catalog to this point because the analyses of the two
are interrelated.  

\subsection{{Completeness of Revised Relative to Original NLTT}
\label{sec:comprnltt}}

\begin{figure}
\epsscale{0.6}
\plotone{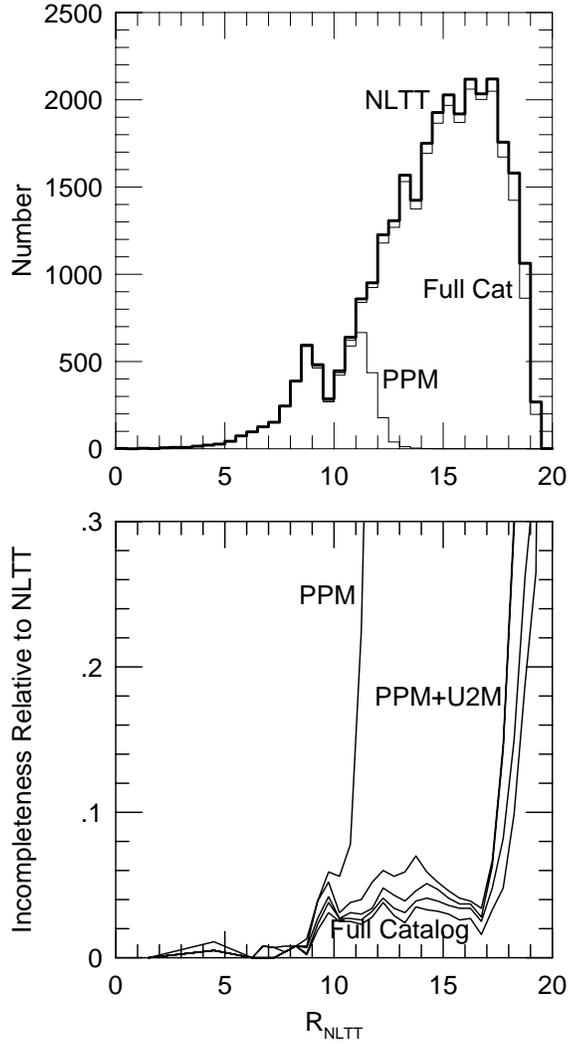}
\caption{\label{fig:completeRnltt}
Top panel: number of catalog stars as a function of $R_\nltt$.  The upper
histogram shows the total number in NLTT.  The lower histogram shows the number
of these recovered by our bright-end search of PPM catalogs (Paper I).
The middle historam (very close to the upper one) 
shows all stars in our catalog.
The bump at $V\sim 9$ is an artifact of NLTT photometry.
Bottom Panel:  close up view of incompleteness relative to NLTT for various
subsets of our catalog, as a function of $R_\nltt$.  The subsets progressively
add the PPM search, the USNO/2MASS search, common proper motion companions
of other stars in the catalog, stars identified only in 2MASS, and stars
identified only in USNO.  The last of these is equivalent to our full
catalog.  The catalog recovers about 97\% of NLTT stars to $R_\nltt\sim 17$.
}\end{figure}

	Figure \ref{fig:completeRnltt}a shows the total number of NLTT
stars as a function of $R_\nltt$ together with the total number of these
in our catalog found by 1) PPM searches alone, 2) all techniques.  
The ``bump'' and ``dip'' at $R_\nltt\sim 9-10$ is an
artifact of the NLTT photometry system and does not appear in other
passbands.  Figure \ref{fig:completeRnltt}b gives the fractional 
incompleteness, and breaks down our catalog entries into five cumulative
classes: 1) PPM only, 2) ... + USNO/2MASS, 3) ... + CPM, 4) ... + 2MASS-only,
5) ... + USNO-only.  For purposes of this plot, we consider only stars in
2MASS-covered areas.  These are defined as stars that have an identified 2MASS
counterpart, or for USNO-only stars, whose known position lies in a
2MASS area, or for NLTT-only stars, whose NLTT position lies in a 2MASS area.
The figures show: first that our PPM search is seemlessly integrated with
our faint searches;  second, that we recover about 97\% of NLTT stars down
to $R_\nltt=17$ and 95\% at $R_\nltt=18$, beyond which NLTT itself is becoming
quite incomplete; third that the auxiliary searches 
(CPM, 2MASS-only, USNO-only) account for only a few percent of our detections
for $10\la R_\nltt \la 17$, but then the fraction rises to about one half.

\subsection{{Relative Completeness of NLTT vs.\ Galactic Latitude}
\label{sec:homosinb}}

	To further investigate problems of completeness, we must first
establish over what range of Galactic latitudes the underlying NLTT
survey is homogeneous.  That is, whatever level of completeness it
actually achieves as a function of magnitude, how uniform is its coverage
over the sky?  We showed in Paper I by direct comparison with Hipparcos and 
Tycho-2, that at bright $(V\la 11)$ magnitudes NLTT is close to 100\% 
complete for $|b|\ga 15^\circ$, but that its completeness falls to 75\%
close to the plane, even for these bright stars.  For the faint stars we
do not have an independent compilation of proper motions, so we cannot
directly establish the absolute completeness of the catalog.  (We
will carry out an indirect absolute measurement in a forthcoming paper.)
However, by making use of both the RPM discriminator $\eta$ 
(eq.\ [\ref{eqn:eta}]) and the much greater dynamic range of the full
catalog presented here, we can give a much more detailed picture of
the relative completeness as a function of various variables.

\begin{figure}
\epsscale{0.7}
\plotone{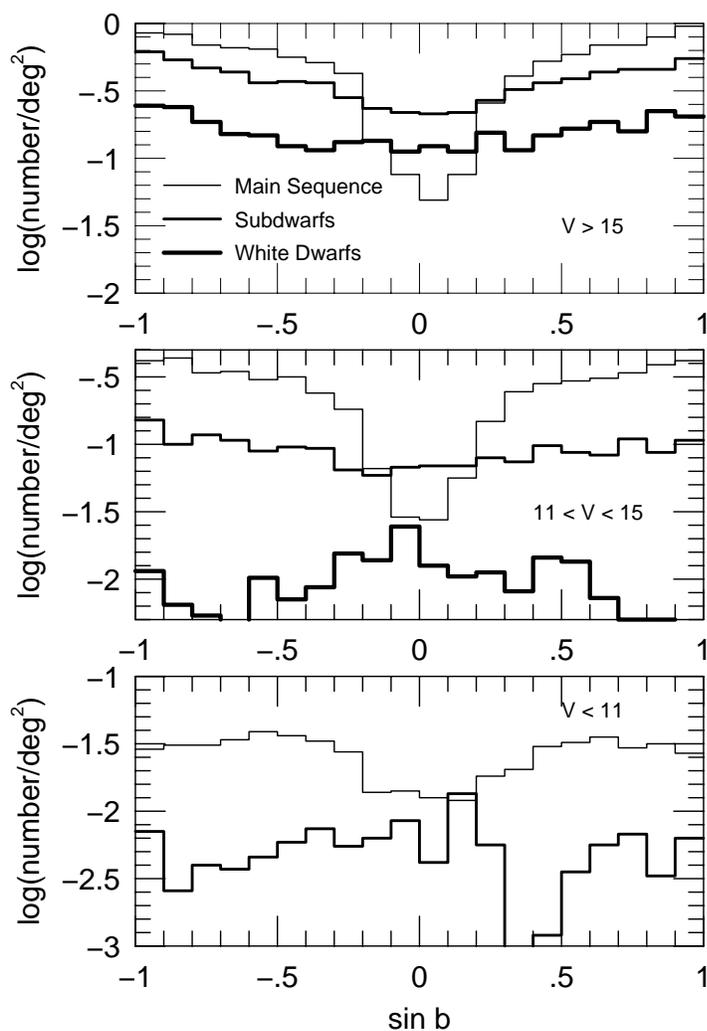}
\caption{\label{fig:rmplatcomp}
Surface density of stars in our catalog as a function of Galactic latitude
for three different stellar populations and three different magnitude bins.
Only stars with $J$ photometry (needed for star classification with the RPM
diagram) are included.  Main sequence stars show a drop of a factor $\sim 10$
in density close to the plane, but subdwarfs and white dwarfs do not.  The
tendency for the counts of all populations to rise at high latitudes in the
faintest bin compared to the flat behavior of the subdwarfs in the mid bin
and the main sequence stars in the bright bin is easily explained from simple
kinematic arguments.
}\end{figure}

	Figure \ref{fig:rmplatcomp} shows the number of WDs, SDs, and
MS stars per square degree as a function of Galactic
latitude, for three different magnitude ranges.  We classify the stars
using equations (\ref{eqn:eta}) and (\ref{eqn:discrim}), which of
course means that stars lacking $J$ photometry will not appear in these
plots.  To determine the surface density (stars deg$^{-2}$), we
calculate the fraction of each latitude bin that is covered by the 
2MASS release and that lies to the north of our cutoff in declination.
(To allow for the somewhat ragged boundary of USNO-A1/POSS I coverage
near $\delta\sim 33^\circ$, we set this cutoff at 
$\delta>-32.\hskip-2pt ^\circ 4$.)\ \  However, we do not attempt to
compensate for stars that fail to appear in 2MASS due either to saturation,
faintness, or other causes.  For $10 \la V \la 17.5$ the fraction of these
is only $\sim 1\%$. On the other hand, bright stars $V\la 6$ are virtually
absent from these plots due to saturation, and saturation remains to some
detectable degree until $V\sim 10$.  This effect should not influence the
{\it relative} number of stars as a function of Galactic latitude, which
is the main point of Figure \ref{fig:rmplatcomp}.  In any event, completeness
questions at the bright end are more effectively investigated by the techniques
of Paper I (see especially, Fig.\ 5).  They are included here mainly as a 
point of reference.   The very faintest blue stars from NLTT will be missing 
from these plots because of the 2MASS flux threshold.

	The results shown in Figure \ref{fig:rmplatcomp} are quite unexpected.
One often hears of the severe incompleteness of NLTT close to the Galactic
plane, but the real story is more complex:  NLTT {\it is} substantially
less complete in the plane, but only for MS stars.  By contrast, NLTT coverage
of SDs is near uniform over the sky and its coverage of WDs appears to be
completely uniform.  For MS stars, there is an evident dropoff
in counts over the interval $-0.2<\sin b<0.3$ in all three magnitude ranges.
It is not very pronounced for bright stars, but one already knows from Paper I
that NLTT is more than 85\% complete averaged over this range.  However,
it is quite pronounced in the other two ranges, falling by a factor 
$\sim 10$ over only $\sim 15^\circ$ in each case.  One also notices a 
more gradual decline in the MS density going from high to low latitudes.
For example, in the faintest bin, it falls by a factor $\sim 2$ between the 
poles and $b\sim \pm 15^\circ$.  A priori, one does not know if this is
due to a genuine change in density or to an extension of the obvious
imcompleteness near the plane to higher latitudes.  In our further 
analysis, we will adopt the former explanation for two reasons.  First,
one does expect a general trend of this sort because a large fraction
of the fainter MS stars entering a proper-motion limited sample are
from the old disk.  These have on average substantially higher transverse 
speeds seen towards the poles than in the plane because of asymmetric drift,
and are therefore selected over a larger volume.  Second, as we now discuss,
this trend is also seen in SDs where it is expected on similar grounds,
but this time without the problem of ``contamination'' from a slower
population.

	There are too few SDs in the bright bin to draw any meaningful
conclusion.  In the middle bin $11<V<15$, the SD distribution appears
to be almost completely flat, in very striking contrast to the MS
distribution in the same magnitude range.  In particular, near the plane
where the MS density suffers a factor 10 decline, the SD distribution is
virtually flat and if it declines, does so by at most a few tens of percent.
The fact that the MS counts gradually rise toward the poles while the SD
counts do not
is easily explained.  For a population of characteristic transverse speed 
$v_\perp$, a survey will saturate its proper-motion limit $\mu_\lim$ only
for stars with $M_V$ brigher than,
\begin{equation}
M_V < V - 5\log{v_\perp\over 10\,\pc\,\mu_\lim}
\label{eqn:absmag}
\end{equation}
For SDs with $v_\perp\sim 300\,\kms$ surveyed at $V=15$ to the NLTT limit
$\mu_\lim=180\,\masyr$, this corresponds to $M_V<7.3$.  There are very few
subdwarfs at these bright magnitudes that could ``take advantage'' of the
higher transverse velocities seen toward the poles.  On the other hand, for
typical MS speeds,  $v_\perp\sim 50\,\kms$, the limit is $M_V<11.1$.  Since
this is close to the peak of the MS luminosity function (LF), a large fraction
of MS stars are seen more readily toward the poles than the plane.  The
same effect explains the patterns seen in the faintest bin which, as we will
see in \S\ \ref{sec:sghighlat},
 contains a substantial number of stars to $V=18$.  The limit
imposed by equation (\ref{eqn:absmag}) for subdwarfs at this magnitude is
$M_V<10.3$, which includes a large fraction of the SD LF.  Similarly, we
expect the effect to be even stronger for MS stars in the faintest bin
than the middle bin, because now the whole peak of the MS LF is included.
On the other hand, in the brightest bin, the limit for MS stars is $M_V<7.1$
which includes a very small fraction of the MS LF.  We therefore
expect the slope to be
small, and it is.

	There are significant numbers of WDs only in the faintest
bin.  Since WDs in NLTT mostly have the same kinematics as the MS, the
limit imposed by equation (\ref{eqn:absmag}) is $M_V<14.1$.  Many WDs satisfy
this constraint, so we expect to find more WDs near the poles, which is
actually the case.  Note that in the faintest bin, neither the WDs nor
the SDs show any significant tendency to drop off close to the plane.

	Thus, a consistent picture emerges from Figure \ref{fig:rmplatcomp}:
MS completeness is very severely affected by proximity to the plane but
SD and WD completeness are barely affected at all.  The most plausible
explanation for this is that while Galactic-plane fields are extremely
crowded and therefore in general subject to confusion, {\it blue} stars 
are no more common in the plane than anywhere else.  Hence, by focusing
on blue objects in the plane, Luyten was able to recover most high-proper
motion SDs and WDs, even while he lost the overwhelming majority ($\sim 90\%$)
of the MS stars.

\subsection{{Properties of Revised NLTT at High Latitude}
\label{sec:sghighlat}}

\begin{figure}
\epsscale{0.75}
\plotone{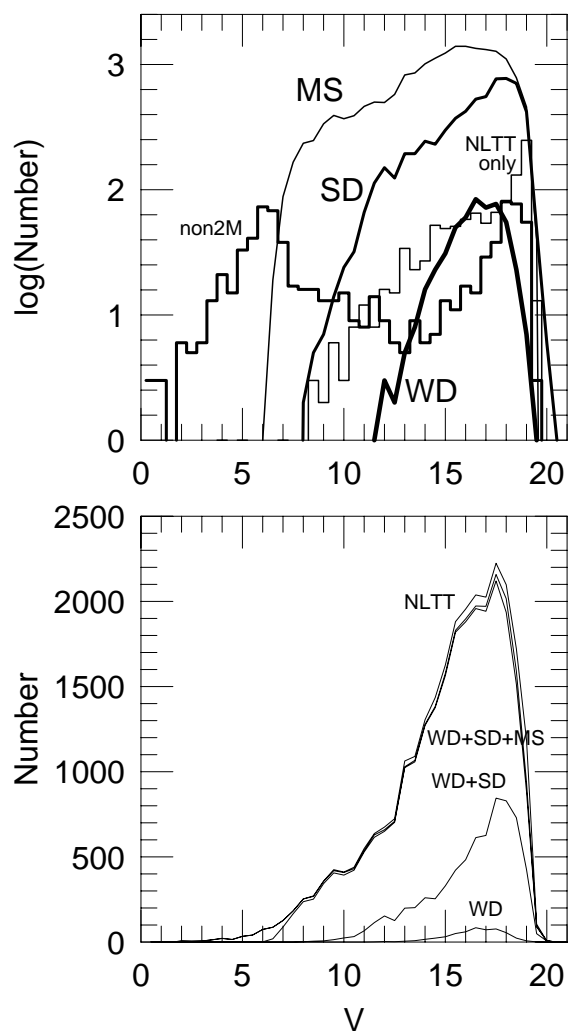}
\caption{\label{fig:vdis}
Distributions of main sequence (MS) stars , subdwarfs (SDs) and white dwarfs
(WDs) as a function of apparent magnitude $V$.  Upper panel: the curves show
the stars in our catalog having $J$ photometry (so they can be classified).
The solid histogram shows NLTT stars that we failed to recover and the bold
histogram stars that are in our catalog but do not have $J$ 
photometry.  Bottom Panel:  successive
cumulative distributions of data from top panel, including first only
WDs, then adding successively SDs, MS, non2M, remainder, so that the highest
curve represents all NLTT stars.  Stars are restricted to 2MASS areas,
$\delta>-32.\hskip-2pt ^\circ 4$, and $\sin b<-0.2$ or $\sin b > 0.3$.
}\end{figure}

	We now investigate the properties of our catalog over the regions
of the sky that are not affected by the severe incompleteness of MS stars
near the plane.  In Figure \ref{fig:vdis}a, we show the (logarithm of the)
number of WDs, SDs and MSs as a function of $V$.  To minimize the problems
of incompleteness near the plane, we have restricted these counts to regions
$\sin b< -0.2$ or $\sin b > 0.3$.  As before, we include only areas covered
by the 2MASS release and $\delta>-32.\hskip-2pt ^\circ 4$.  These restrictions
apply to 38.0\% of the sky.  Also shown is a bold histogram of the stars
in our catalog that lie in this 38\% of sky but were absent from 2MASS
(possibly due to saturation),
as well as a solid histogram of NLTT stars in this region that we failed
to recover.  Figure \ref{fig:vdis}b shows the cumulative number
at each magnitude, progressively including WDs, SDs, MS stars, stars
without 2MASS data, and NLTT stars missing from our catalog.  
The uppermost curve
(all NLTT stars) turns over sharply at $V=17.5$, which would seem to
indicate the onset of substantial incompleteness at this magnitude.
However, the upper panel shows that this
assessment ignores significant subtleties.

	First note that the MS curve flattens and turns over at $V\sim 15$, 
as does the WD curve, but that the SD curve continues to rise to 
$V\sim 17.5$.
A plausible explanation for these divergent behaviors is that completeness
does not begin to drop off rapidly until past $V\sim 17.5$, but that the MS
and WD counts are adversely impacted by other effects.  At faint magnitudes,
the former are just running out of stars.  Recall from equation 
(\ref{eqn:absmag}) that at the proper motion limit $\mu_\lim=180\,\masyr$,
MS stars with transverse speeds $v_\perp\sim 50\,\kms$ have absolute 
magnitudes $M_V\sim V-3.8$.  Hence, for $V\ga 15$ the survey is probing the LF
beyond its peak where there are a declining number of stars. 
We suspect that the
WD are becoming too faint to show up in 2MASS and so they are not being
classified as such by our optical/IR discriminant, $\eta$.  At fainter 
magnitudes, they are too faint even for USNO, and so are not being recovered
by our catalog at all.

	To test this idea, we classify USNO-only stars using the 
$(B-R)_\usno$ RPM diagram.  We have previously shown (\citealt{rpm}, see 
Fig.\ 4) that
this diagram discriminates quite well between WDs and other stars, but does not
discriminate effectively between SDs and MS stars.  Based on the USNO RPM
diagram, we classify USNO-only stars as WDs if they satisfy
\begin{equation}
 R_\usno -3.5(B-R)_\usno + 5\log \mu > 10, \qquad \rm (USNO\ WDs).
\label{eqn:usnowd}
\end{equation}
The NLTT RPM can also be used to classify WDs \citep{rpm}.
Although on a star-for-star basis, this classification is less reliable
than the one base on the USNO RPM diagram, and substantially less reliable than
the one based on the $(V-J)$ RPM diagram that we mostly use here, it should
be adequate to address the statistical questions posed in this section.
Using Figure 3 from \citet{rpm}, we derive,
\begin{equation}
R_\nltt - 7(B-R)_\nltt + 5\log \mu > 10, \qquad \rm (NLTT\ WDs).
\label{eqn:nlttwd}
\end{equation}

\begin{figure}
\epsscale{1.0}
\plotone{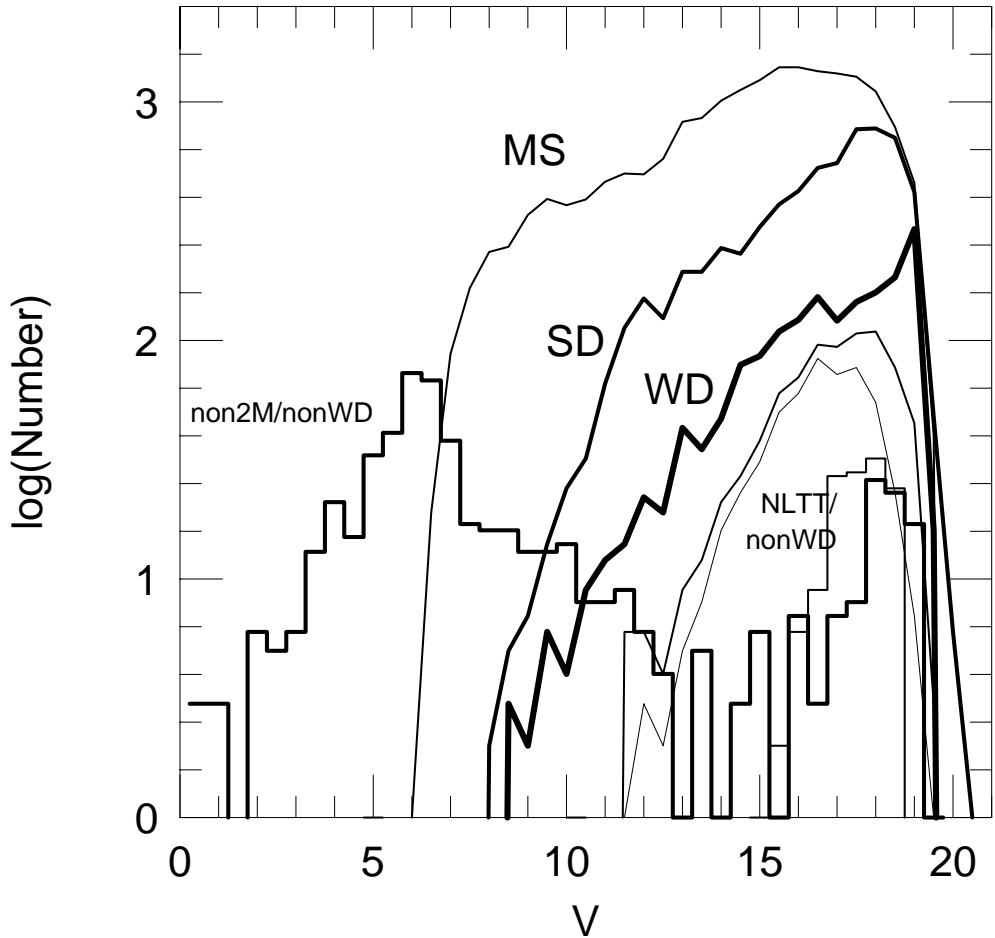}
\caption{\label{fig:vdis2}
Similar to Fig.\ \ref{fig:vdis}a except that now the non2M and NLTT
histograms from that figure have been broken down into their WD and
nonWD components.  The nonWD components remain as histograms.  The
bold WD curve now includes the WDs from all three sources.  The lower of the
two unlabeled solid curves is the same as the WD curve from 
Fig. \ref{fig:vdis}a, and the upper one is the sum of this and the USNO WDs.
Notice that the WD curve now rises all the way to $V=19$ indicating that
probably the leveling of the MS and SD curves is due to a genuine dearth
of stars rather than incompleteness.
}\end{figure}

	Figure \ref{fig:vdis2} is similar to Figure \ref{fig:vdis}a, except
that it reflects the reclassification of the USNO-only and NLTT-only stars.
The bold curve now shows  the total of all the WDs, while the histograms
show the non2MASS/nonWDs and NLTT-only/nonWDs.  Also shown are two curves that
give the $(V-J)$ RPM-classified WDs (same as WD curve in 
Fig.\ \ref{fig:vdis}a), and the sum of this curve and the USNO-only WDs.
This diagram has several important features.  First, it shows that the
WD counts continue rising to $V=19$, actually slight past the end of the
SD rise.  Hence, it seems likely that the SD turnover at $V\sim 17.5$
is due to a real decline in the LF at faint magnitudes, just as was the
case for the MS.  Second, while we do not know what the unclassified
USNO-only and NLTT-only stars are at the faint end, their combined counts
are a factor 10 below the SD counts, and even farther below the MS counts.
Hence, however they are classified, they cannot alter the qualitative
picture presented by this diagram.

	These results imply that our catalog can be used to study the
statistical properties of a large sample of subdwarfs down to $V\sim 18$.
We have initiated such a study and will report our results soon.

\section{{Going South?}
\label{sec:south}}

	Once the full 2MASS catalog is released, it will be relatively
straightforward to extend our catalog to the other 33\% of the sky
that lies north of $\delta=-33^\circ$.  However, the prospects for extending
it to the south are less promising.  First, as we discussed in 
\S\ \ref{sec:scope}, the great majority of NLTT stars are missing from
USNO-A.  In the north, even when a NLTT star was missing from USNO-A, 
we were frequently able to recover it by looking for the 2MASS counterpart
at the position predicted by assuming that the star was inside the NLTT
rectangle and that the NLTT proper motion was approximately correct.  See \S\
\ref{sec:2mass_only}.  However, the reason that this method was effective
is that over most of the northern sky, the star does actually lie in the
rectangle about 80\% of the time.  From Figure \ref{fig:rectlat}, one
sees that progressively fewer stars are in the rectangle in the south,
and virtually none are for $\delta<-45^\circ$.  Moreover, a 2MASS-only 
identification yields a position and infrared photometry, which are useful
for many applications, but not a proper motion. At present only 13\%
of the area $\delta < -33^\circ$ is covered by the 2MASS release, but
the above factors lead us to believe that after full release, our technique
will not be very effective in this region.  In any event, NLTT is generally
limited to $V\la 15$ in the South and the brighter stars among these are
already covered by our PPM search.

\section{{Conclusion}
\label{conclude}}

	The catalog presented here gives improved astrometry and photometry
for the great majority of stars in the NLTT that lie in the overlap of
the areas covered by the second incremental 
2MASS release and those covered by POSS I (basically
$\delta>-33^\circ$).  In addition, essentially all bright NLTT stars over the
whole sky have been located in PPM catalogs and, whenever possible, the
close binaries among them have been resolved using TDSC.  We
recover essentially 100\% of NLTT stars $V<10$, about 97\% for
$10<V<18$, and a declining fraction thereafter.  Most of the faint stars
that we do not recover are WDs, which are too faint and too blue to
show up in one or both of USNO-A or 2MASS, the two catalogs that we combine
to obtain our astrometry and photometry.  The catalog contains a total of
36,020 entries for 35,662 NLTT stars including a total of 723 entries for 361 
NLTT ``stars'' that have been resolved in TDSC.  Of these 36,020 entries,
1353 ($\sim 4\%$) contain new positions and photometry but not new
proper motions because they have been identified in only USNO-A or 2MASS,
but not both.

	The new positions are accurate to 130 mas. The new proper motions
are accurate to $5.5\,\masyr$, more than a 3-fold improvement over NLTT.
Narrow angle proper motions are accurate to $3\,\masyr$.  The catalog provides
a powerful powerful means to investigate SDs, WDs, faint MS stars as well
as CPM binaries.


\acknowledgments We thank S.\ Frink for providing access to the Starnet
catalog, which has been made available to the SIMBAD database
by S.\ R{\" o}ser but is not yet accessible there. We acknowledge the use
of CD-ROM versions of USNO-A1.0 and A2.0 catalogs provided to us by D.\
Monet.
This publication makes use of catalogs from the Astronomical Data
Center at NASA Goddard Space Flight Center, VizieR and SIMBAD databases
operated at CDS, Strasbourg, France, and data products from the Two
Micron All Sky Survey, which is a
joint project of the University of Massachusetts and the IPAC/Caltech,
funded by the NASA and
the NSF.  This work was supported by JPL contract 1226901.

\clearpage

\clearpage
\begin{deluxetable}{l r r r r r}
\tabletypesize{\footnotesize}
\tablecaption{Number of USNO/2MASS matches \label{tab:usno2mass}}
\tablewidth{0pt}
\tablehead{

\colhead{} &
\multicolumn{2}{c}{rectangles} &
\multicolumn{2}{c}{circles} &
\colhead{total} \\
\colhead{} &
\colhead{unique}   &
\colhead{resolved}   &
\colhead{unique}   &
\colhead{resolved}   &
\colhead{} \\
}
\startdata
USNO-A1 &  2426 &   78 &  517 &  174 &  3195 \\
USNO-A2 & 15404 &  534 & 3574 &  974 & 20486 \\
total &   17830 &  612 & 4091 & 1148 & 23681 \\
\enddata
\end{deluxetable}

\end{document}